\documentclass[a4paper,11pt]{article}
\usepackage[compat=1.1.0]{tikz-feynman}
\usepackage{jheppub} 
\usepackage[british]{babel}
\renewcommand{\acknowledgments}{\section*{Acknowledgements}}
\usepackage{lineno}
\usepackage{feynmp-auto}
\usepackage{mathabx}
\usepackage{nicematrix}
\usepackage{feyn}
\usepackage{bigints}
\usepackage{algorithm}
\usepackage{algpseudocode}
\usepackage{hyperref}
\makeatletter
\gdef\@fpheader{}
\makeatother

\usetikzlibrary{
    decorations.markings,
    decorations.pathmorphing,
    decorations.text,
}

\usepackage{xcolor}
\usepackage{listings}

\lstdefinestyle{terminal}{
    language=bash,
    basicstyle=\small\ttfamily,
    backgroundcolor=\color{black!5},
    frame=single,
    rulecolor=\color{black!30},
    breaklines=true,
    columns=fullflexible,
    keepspaces=true,
    showstringspaces=false
}

\title{\boldmath \texttt{Pyresias}: How to write a parton shower}

\preprint{MCNET-24-08}

\author[1]{Andreas Papaefstathiou}
\affiliation[1]{Department of Physics, Kennesaw State University, 830 Polytechnic Lane, Marietta, GA 30060, USA}

\emailAdd{apapaefs@kennesaw.edu}

\abstract{We present a hands-on introduction to the construction of a
  parton shower for final-state radiation in electron--positron
  annihilation. The tutorial is intended for students with a basic
  knowledge of perturbative quantum chromodynamics and
  \texttt{Python}. Starting from collinear factorization, we derive the Sudakov
  no-emission probability and explain how the veto algorithm generates
  successive parton branchings. We then introduce colour coherence through angular ordering and reconstruct the final-state momenta, including the recoil needed to conserve energy and momentum. The accompanying \texttt{Python} program, \texttt{Pyresias}, develops these ingredients in stages: a single quark line, a quark--antiquark event, and a branching tree containing quark and gluon splittings. Explicit sampling formulae, pseudocode, and exercises connect the derivations to the implementation. We test the veto algorithm against the numerically integrated no-emission probability and first-emission distribution, and compare the reconstructed events with the angular-ordered shower of \texttt{Herwig 7}, using matching parton-level settings. These comparisons provide reproducible implementation checks within the stated approximations. The tutorial also identifies how choices of ordering variable, recoil, and branching channels connect to precision shower development.}

\begin{document}
\maketitle
\raggedbottom

\section{Introduction}
\label{sec:intro}

Perturbative calculations in quantum chromodynamics (QCD) can be difficult. In particular, including successively higher orders in the perturbative expansion in the strong coupling constant ($\alpha_S$) becomes increasingly challenging. For sufficiently inclusive quantities, such as the total cross section, perturbative predictions are well-behaved power series in $\alpha_S$, with coefficients that are $\mathcal{O}(1)$. For many quantities of interest, however, these coefficients may contain large logarithms that have to be `resummed' to all orders to obtain reliable predictions. Parton showers provide a means of achieving this resummation by including enhanced terms associated with `soft' (low-energy) and `collinear' (small-angle) radiation to all orders.\footnote{See, e.g.~\cite{Fox:1979ag,Marchesini:1983bm,Giele:1990vh,Webber:1992nx}. For recent reviews in the context of particle physics event generators, see, e.g.~\cite{Buckley:2011ms,Campbell:2022qmc}.} The parton-shower formalism is a central component of Monte Carlo event generators for particle colliders and can be combined with non-perturbative models that describe the conversion of partons (i.e.\ quarks and gluons) into hadrons via the process known as `hadronization'.

The purpose of the present article is to connect the physics of QCD radiation to a parton shower that the reader can build, run, and modify. The accompanying \texttt{Python} program, \texttt{Pyresias}, and its JupyterLab notebook are publicly available~\cite{repo}. We address students who have encountered tree-level amplitudes, cross sections, and elementary QCD, and who can write basic \texttt{Python}. Familiarity with parton showers is not assumed. The intended outcome is that the reader can explain a no-emission probability, implement a veto sampler, construct a branching history, and test the reconstructed event.

The material builds on the textbook by Ellis, Stirling, and
Webber~\cite{Ellis:1996mzs}, the lecture notes by
Gieseke~\cite{stefan}, the treatment of veto algorithms by
L\"onnblad~\cite{Lonnblad:2012hz}, and the \texttt{Herwig}
manuals~\cite{Bahr:2008pv,Bellm:2025pcw}. H\"oche provides a
complementary introduction to parton-shower event
generators~\cite{Hoche:2014rga}, and the tutorial of H\"oche and
Prestel~\cite{hocheprestel} builds a transverse-momentum-ordered
shower in \texttt{Python}, based on Catani--Seymour dipole kernels
with emitter--spectator kinematics. The present article is
complementary to that tutorial: it constructs an angular-ordered
shower with global recoil, develops it from a single quark line to
complete events with all three QCD branching channels, and compares
the result with \texttt{Herwig} under matching settings. The
analytical results have been long established, and the emphasis is on making their implementation and limitations explicit.

We choose angular ordering because it exposes the connection between
soft-gluon interference and a probabilistic branching
algorithm~\cite{Marchesini:1983bm}. It also permits a controlled
comparison with the angular-ordered shower of \texttt{Herwig 7}, which
remains its default shower and is available alongside a dipole
shower~\cite{Platzer:2009jq,Bellm:2015jjp,Bellm:2025pcw}. Other useful
constructions include the dipole cascade~\cite{Gustafson:1987rq},
transverse-momentum-ordered showers with dipole
recoil~\cite{Sjostrand:2004ef}, and showers based on Catani--Seymour
dipole factorization~\cite{Schumann:2007mg}. The
\texttt{Pythia}~\cite{Bierlich:2022pfr} and
\texttt{Sherpa}~\cite{Sherpa:2024mfk} event generators provide further
examples. The antenna shower \texttt{Vincia}~\cite{Fischer:2016vfv}
and \texttt{Dire}, which combines dipole-cascade ideas with collinear
parton evolution~\cite{Hoche:2015sya}, illustrate further choices of
radiation kernels and kinematics. The \texttt{PanScales} project develops parton showers with
systematically improved logarithmic accuracy~\cite{Dasgupta:2020fwr,vanBeekveld:2024wws},
and provides a public framework for their implementation and
validation~\cite{vanBeekveld:2023ivn}. The
\texttt{Alaric}~\cite{Herren:2022jej} and
\texttt{Apollo}~\cite{Preuss:2024vyu} showers also combine coherent
radiation with global recoil to achieve consistency with
next-to-leading-logarithmic resummation. Ordering, radiation kernels, and recoil are related design
choices, and the ordering variable alone does not specify a shower's accuracy.

The tutorial proceeds in three stages. We first follow only
$q\rightarrow qg$ branchings along a massless quark line, leaving
emitted gluons unevolved. We then shower both legs of
$e^+e^-\rightarrow q\bar q$ and reconstruct their momenta. Finally, we
add $g\rightarrow gg$ and $g\rightarrow q\bar q$ and evolve both
daughters after every branching. The initial restriction isolates the
Sudakov construction and is therefore not intended to provide a complete leading-logarithmic QCD shower. Throughout, we work at parton level in the centre-of-mass frame and omit hadronization, spin correlations, and matrix-element corrections.

Section~\ref{sec:innerworkings} develops the branching probabilities, veto algorithm, and coherence argument. Section~\ref{sec:shower} gives the implementation, from a quark line to a branching tree, followed by numerical checks and exercises. Section~\ref{sec:conclusions} presents the conclusions. Appendix~\ref{app:angular-distribution} derives the exact single-emission angular distribution. Appendix~\ref{app:sudakovproof} proves the veto algorithm, and appendix~\ref{app:rotations} describes the rotations used in momentum reconstruction.

\section{The Inner Workings of a Parton Shower}\label{sec:innerworkings}

\subsection{Single QCD Emissions in Electron--Positron Annihilation}

\begin{figure}[!htp]
\centering


\includegraphics[]{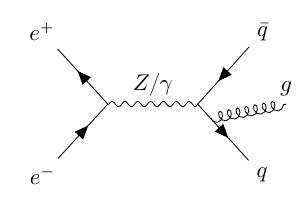}
\caption{Example Feynman diagram for electron--positron annihilation into a quark--antiquark pair with an emitted gluon.}
\label{fig:eeqqbg}
\end{figure}

We begin by considering electron--positron annihilation in the centre-of-mass frame, i.e.\ $e^+ e^- \rightarrow q\bar{q}g$. An example Feynman diagram contributing to this process is shown in fig.~\ref{fig:eeqqbg}. We define the centre-of-mass energy as $Q$, such that the four-vector of the intermediate (off-shell) gauge boson ($Z$ boson or photon, $\gamma$) is given by $q^\mu = (Q, \mathbf{0})$. We then define the momentum fractions of the three outgoing particles as $x_i = 2 E_i / Q$, where $x_i\in [0,1]$ and $i=1,2,3$, with $3$ being the index of the emitted gluon. This implies that $\sum_{i=1}^{3} x_i = 2$.

Evaluating the Feynman diagrams gives the double-differential cross section with respect to the quark momentum fractions (see, e.g.~\cite{Field:1989uq}):
\begin{equation}\label{eq:ee-energy-distribution}
\frac{ \mathrm{d} \sigma }{ \mathrm{d} x_1 \mathrm{d} x_2} = \sigma_0 \frac{ C_F \alpha_S }{ 2\pi} \frac{ x_1^2 + x_2^2 }{(1-x_1) (1-x_2)}\;,
\end{equation}
where $C_F=4/3$ is the quark colour factor, $\alpha_S$ is the QCD coupling constant, and $\sigma_0$ is the leading-order cross section (i.e.\ the total cross section for $e^+e^- \rightarrow q\bar{q}$).
This differential cross section contains two kinds of singularities:
\begin{itemize}
    \item collinear singularities as $x_1 \rightarrow 1$ or $x_2 \rightarrow 1$;
    \item soft singularities as both $x_1, x_2 \rightarrow 1$ at the same time.
\end{itemize}
In the singular limits, the angular dependence can be written as
\begin{equation}\label{eq:ee-angular-singular}
    \frac{ \mathrm{d} \sigma }{ \mathrm{d} x_3 \mathrm{d} \cos \theta} = \sigma_0 \frac{ C_F \alpha_S }{ 2\pi} \left[ \frac{2}{\sin^2 \theta} \frac{ 1 + (1-x_3)^2}{x_3}+ \mathrm{non-singular~terms} \right]\;,
\end{equation}

where $\theta = \angle(q,g)$. The exact change of variables, including the non-singular term, is derived in appendix~\ref{app:angular-distribution}. The decomposition
\begin{equation}
\frac{2}{1-\cos^2\theta}
 = \frac{1}{1-\cos\theta}+\frac{1}{1+\cos\theta}
\end{equation}
separates the two collinear endpoints. Near the quark direction,
$1-\cos\theta\simeq\theta^2/2$, so the positive angular integration measure is
$|\mathrm{d}\cos\theta|/(1-\cos\theta)\simeq\mathrm{d}\theta^2/\theta^2$.
The other endpoint gives the corresponding measure for
$\bar\theta=\angle(\bar q,g)$ near the antiquark direction. These statements
refer to the collinear limits, where the other hard parton recoils against
the unresolved pair.

\begin{figure}[!htp]
\centering
\includegraphics[]{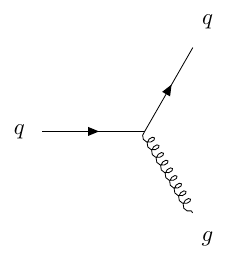}
\caption{The splitting $q\rightarrow qg$ is represented by the universal splitting function of eq.~\ref{eq:splqgq}.}
\label{fig:qgq}
\end{figure}
This allows us to identify two `jets' and rewrite the differential cross section as
\begin{equation}\label{eq:singleem}
\mathrm{d} \sigma \simeq \sigma_0 \sum_{\mathrm{jets}}  \frac{\mathrm{d} \theta^2} { \theta^2 } \mathrm{d} z\frac{ \alpha_S } { 2\pi } C_F \frac{ 1 + z^2 } {1- z} \;,
\end{equation}
where $z$ is the fraction retained by the quark in its collinear branching, so that $x_3\simeq 1-z$ in this limit. We can also define the splitting function for a quark emitting a gluon (i.e.\ $q\rightarrow qg$) as
\begin{equation}\label{eq:splqgq}
P(z) =  C_F \frac{ 1 + z^2 } {1-z} \;.
\end{equation}

This is the real-emission quark splitting kernel~\cite{Altarelli:1977zs}. Its collinear factorization is independent of the hard process. In an inclusive evolution equation, virtual contributions also appear through endpoint terms; here they will enter through the Sudakov factor. The index notation for splitting functions differs between references, so we specify the branching $q\rightarrow qg$ and the meaning of $z$ explicitly.

At fixed $z$, the parent virtuality $m_a^2$ and relative transverse momentum $p_T^2$ are proportional to the opening angle squared in the collinear limit. Hence the logarithmic measures obey
\begin{equation}
    \frac{\mathrm{d}\theta^2}{\theta^2}
    =\frac{\mathrm{d}m_a^2}{m_a^2}
    =\frac{\mathrm{d}p_T^2}{p_T^2}
    =\frac{\mathrm{d}t}{t}
    \qquad\text{(fixed $z$, collinear limit)}\;,
\end{equation}
where $t$ denotes a squared evolution scale. This equality of single-emission measures does not make the ordering prescriptions equivalent for several emissions.


We also need to introduce a resolution parameter, $t_0$, which could be, e.g.\ a cut-off in $p_T$. This prevents us from reaching the $\theta \rightarrow 0$ singularity. Emissions below $t_0$ are unresolvable. Their contributions combine with virtual corrections to give a finite result. We account for these corrections through unitarity, using the Sudakov form factor introduced below. Schematically:
\begin{equation*}
  \begin{tikzpicture}[baseline=-\the\dimexpr\fontdimen22\textfont2\relax]
    \begin{feynman}[inline=(a)]
    \vertex (a) {};
    \vertex [right=of a] (b);
    \vertex [above right=of b] (d);
    \vertex [right=of b] (c);
    \diagram*[small] {
       (a) -- [fermion] (b) -- [fermion] (c) -- [fermion] (c),
       (b) -- [gluon] (d),
    };
    \end{feynman}
  \end{tikzpicture} ~~+
    \begin{tikzpicture}[baseline=-\the\dimexpr\fontdimen22\textfont2\relax]
    \begin{feynman}[inline=(a)]
    \vertex (a) {};
    \vertex [right=of a] (b);
    \vertex [right=of b] (c);
    \vertex [right=of c] (d);
    \vertex [right=of d] (e);
    \diagram*[small] {
       (a) -- [fermion] (b) -- [fermion] (c) -- [fermion] (d),
       (b) -- [gluon, half left] (c),
    };
    \end{feynman}
  \end{tikzpicture}=~\mathrm{finite}\;,
\end{equation*}
where the diagram with the gluon emission denotes unresolvable collinear emissions. This fact is known as the Bloch--Nordsieck/Kinoshita--Lee--Nauenberg (BN/KLN) theorem~\cite{Bloch:1937pw,Kinoshita:1962ur,Lee:1964is}.

\subsection{Multiple Emissions}\label{sec:multiple}

Let us now consider multiple emissions, extending the single-emission result derived via the process $e^+e^-\rightarrow q\bar qg$. To begin with, we allow emissions only from the original quark and antiquark lines, leaving the emitted gluons unevolved. The branchings $g\rightarrow gg$ and $g\rightarrow q\bar q$ are also present in QCD; we omit them here to simplify the argument and include them in section~\ref{sec:competition}. Their omission does not follow from the collinear approximation.

First, we write the real-emission contribution for one additional gluon in $e^+e^-\rightarrow q\bar q$, following eq.~\ref{eq:singleem}, as
\begin{equation}
 \sigma_1^{\mathrm{real}}(t_0)
 =\sigma_0\sum_{\mathrm{jets}}\int_{t_0}^{t_{\mathrm{max}}}\frac{\mathrm{d}t'}{t'}
 \int_{z_-(t')}^{z_+(t')}\mathrm{d}z\,
 \frac{\alpha_S(\mu(t',z))}{2\pi}P(z)\;,
\end{equation}
where $P(z)$ is the splitting function of eq.~\ref{eq:splqgq}, $t_0$ is the cut-off, and $t_{\mathrm{max}}$ is the starting scale. The limits $z_\pm$ arise from kinematics and can depend on both $t'$ and the cut-off; $\mu$ denotes the coupling scale. Since there are two outgoing quark lines, we can write:
\begin{equation}
 \sigma_1^{\mathrm{real}}(t_0)
 =2\sigma_0\int_{t_0}^{t_{\mathrm{max}}}\mathrm{d}t'\,\Gamma(t')\;,
\end{equation}
where
\begin{equation}\label{eq:rate}
 \Gamma(t')\equiv\frac{1}{t'}\int_{z_-(t')}^{z_+(t')}\mathrm{d}z\,
 \frac{\alpha_S(\mu(t',z))}{2\pi}P(z)\;.
\end{equation}
Thus $\Gamma(t')\mathrm{d}t'$ gives the probability of a branching in an infinitesimal interval, provided that the line has reached that scale without branching. The superscript `real' distinguishes the real-emission contribution from an exclusive shower cross section, which must also account for virtual and unresolved emissions.

We will consider multiple emissions that are strongly ordered in the evolution variable, $t_1\gg t_2\gg\cdots\gg t_n$. In this limit, collinear factorization can be applied successively. The shower extends this approximation over the full ordered region, $t_{\mathrm{max}}>t_1>t_2>\cdots>t_n>t_0$. For the counting argument below, we also assume that the rate $\Gamma$ and its available phase space do not change after an emission.

We then define the real-emission contribution with $n$ resolved gluons as:
\begin{equation}
 \sigma_n^{\mathrm{real}}(t_0)=\sigma_0\Omega_n(t_0,t_{\mathrm{max}})\;,
 \qquad n>0\;,
\end{equation}
so that $\Omega_1(t_0,t_{\mathrm{max}})=2\int_{t_0}^{t_{\mathrm{max}}}\mathrm{d}t'\,\Gamma(t')$. Since $\Omega_n=\sigma_n^{\mathrm{real}}/\sigma_0$, we can write, schematically:
\begin{equation}\label{eq:omeganschematic}
  \Omega_n = \frac{\int \mathrm{d} \Phi_n \left|
  \begin{tikzpicture}[baseline=-\the\dimexpr\fontdimen22\textfont2\relax]
    \begin{feynman}[inline=(a)]
    \vertex[blob] (a) {};
    \vertex [above right=of a] (b);
    \vertex [right=of a] (d) [label=above:{\vdots}];
    \vertex [below=of b] (e);
    \vertex [above right=of b] (f) at (0.25,-0.17);
    \vertex [below right=of a] (c);
    \diagram*[small] {
    (a) -- [fermion] {(b), (c)},
    (a) -- [gluon] (d),
    (a) -- [gluon] (e),
    (a) -- [gluon] (f),
    };
    \end{feynman}
  \end{tikzpicture}\right|^2
}{ \left|\begin{tikzpicture}[baseline=-\the\dimexpr\fontdimen22\textfont2\relax]
    \begin{feynman}[inline=(a)]
    \vertex[dot] (a) {};
    \vertex [above right=of a] (b);
    \vertex [below right=of a] (c);
    \diagram*[small] {
    (a) -- [fermion] {(b),(c)}
    };
    \end{feynman}
  \end{tikzpicture}\right|^2}\;.
\end{equation}
Here $\mathrm{d}\Phi_n$ denotes the resolved emission phase space relative to the underlying Born configuration, and the blob denotes the sum of amplitudes retained in this quark-line example. Before expanding the blob, we must consider the interference between diagrams. Even for one gluon, the exact squared amplitude is:
\begin{equation}\label{eq:interference}
 |\mathcal M_{q\rightarrow qg}+\mathcal M_{\bar q \rightarrow \bar q g}|^2
 =|\mathcal M_{q \rightarrow qg} |^2+|\mathcal M_{\bar q \rightarrow
   \bar q g}|^2
  +2\operatorname{Re}(\mathcal M_{q \rightarrow qg} \mathcal M_{\bar q
  \rightarrow \bar q g}^{*})\;.
\end{equation}
In a suitable physical (e.g.\ light-cone) gauge, the leading collinear
singularity along the quark direction comes from $|\mathcal M_{q
  \rightarrow q g}|^2$, and similarly for the antiquark. The
interference term has at most integrable singularities in these
collinear limits~\cite{Nagy:2019pjp}. This allows us to associate the
leading collinear contribution with a branching from one line,
yielding the splitting kernel after the spin and azimuthal averages
used here. Individual diagrams are not separately gauge invariant,
however, and their squares cannot be interpreted as independent
emission probabilities in general. In particular, soft interference at
fixed angles remains important, and we return to it in section~\ref{sec:soft}.

With these qualifications, the diagrams below represent the leading, strongly ordered collinear branching histories. For example, for one emission, we have:
\begin{equation*}
  \Omega_1 \simeq \frac{\int \mathrm{d} \Phi_1 \left[ \left|
  \begin{tikzpicture}[baseline=-\the\dimexpr\fontdimen22\textfont2\relax]
    \begin{feynman}[inline=(a)]
    \vertex[dot] (a) {};
    \vertex [above right=of a] (b);
    \vertex [below right=of a] (c);
    \vertex [below right=of b] (r) at (0.4, 1.0);
    \draw [gluon] ($(a)!0.4!(b)$)  -- (r);
    \diagram*[small] {
    (a) -- [fermion] {(b), (c)},
    };
    \end{feynman}
  \end{tikzpicture}\right|^2
  + \left|
    \begin{tikzpicture}[baseline=-\the\dimexpr\fontdimen22\textfont2\relax]
    \begin{feynman}[inline=(a)]
    \vertex[dot] (a) {};
    \vertex [above right=of a] (b);
    \vertex [below right=of a] (c);
    \vertex [above right=of a] (r) at (0.4,-1.0);
    \draw [gluon] ($(a)!0.4!(c)$)  -- (r);
    \diagram*[small] {
    (a) -- [fermion] {(b), (c)},
    };
    \end{feynman}
  \end{tikzpicture}\right|^2
  \right]
}{ \left|\begin{tikzpicture}[baseline=-\the\dimexpr\fontdimen22\textfont2\relax]
    \begin{feynman}[inline=(a)]
    \vertex[dot] (a) {};
    \vertex [above right=of a] (b);
    \vertex [below right=of a] (c);
    \diagram*[small] {
    (a) -- [fermion] {(b),(c)}
    };
    \end{feynman}
  \end{tikzpicture}\right|^2}\;.
\end{equation*}
and for two emissions, $n=2$:
\begin{equation*}
  \Omega_2 \simeq \frac{\int \mathrm{d} \Phi_2 \left[ \left|
  \begin{tikzpicture}[baseline=-\the\dimexpr\fontdimen22\textfont2\relax]
    \begin{feynman}[inline=(a)]
    \vertex[dot] (a) {};
    \vertex [above right=of a] (b);
    \vertex [below right=of a] (c);
    \vertex [below right=of b] (r) at (0.4, 0.8);
    \vertex [below right=of b] (q) at (0.4, 1.1);
    \draw [gluon] ($(a)!0.2!(b)$)  -- (r);
    \draw [gluon] ($(a)!0.6!(b)$)  -- (q);
    \diagram*[small] {
    (a) -- [fermion] {(b), (c)},
    };
    \end{feynman}
  \end{tikzpicture}\right|^2
  + \left|
    \begin{tikzpicture}[baseline=-\the\dimexpr\fontdimen22\textfont2\relax]
    \begin{feynman}[inline=(a)]
    \vertex[dot] (a) {};
    \vertex [above right=of a] (b);
    \vertex [below right=of a] (c);
    \vertex [above right=of a] (r) at (0.4,-0.8);
    \vertex [above right=of a] (q) at (0.4,-1.1);
    \draw [gluon] ($(a)!0.2!(c)$)  -- (r);
    \draw [gluon] ($(a)!0.6!(c)$)  -- (q);
    \diagram*[small] {
    (a) -- [fermion] {(b), (c)},
    };
    \end{feynman}
  \end{tikzpicture}\right|^2
    + \left|
    \begin{tikzpicture}[baseline=-\the\dimexpr\fontdimen22\textfont2\relax]
    \begin{feynman}[inline=(a)]
    \vertex[dot] (a) {};
    \vertex [above right=of a] (b);
    \vertex [below right=of a] (c);
    \vertex [above right=of a,label=right:2] (r) at (0.4,-1.0);
    \vertex [below right=of b,label=right:1] (q) at (0.4,1.0);
    \draw [gluon] ($(a)!0.4!(c)$)  -- (r);
    \draw [gluon] ($(a)!0.4!(b)$)  -- (q);
    \diagram*[small] {
    (a) -- [fermion] {(b), (c)},
    };
    \end{feynman}
  \end{tikzpicture}\right|^2
      + \left|
    \begin{tikzpicture}[baseline=-\the\dimexpr\fontdimen22\textfont2\relax]
    \begin{feynman}[inline=(a)]
    \vertex[dot] (a) {};
    \vertex [above right=of a] (b);
    \vertex [below right=of a] (c);
    \vertex [above right=of a,label=right:1] (r) at (0.4,-1.0);
    \vertex [below right=of b,label=right:2] (q) at (0.4,1.0);
    \draw [gluon] ($(a)!0.4!(c)$)  -- (r);
    \draw [gluon] ($(a)!0.4!(b)$)  -- (q);
    \diagram*[small] {
    (a) -- [fermion] {(b), (c)},
    };
    \end{feynman}
  \end{tikzpicture}\right|^2
  \right]
}{ \left|\begin{tikzpicture}[baseline=-\the\dimexpr\fontdimen22\textfont2\relax]
    \begin{feynman}[inline=(a)]
    \vertex[dot] (a) {};
    \vertex [above right=of a] (b);
    \vertex [below right=of a] (c);
    \diagram*[small] {
    (a) -- [fermion] {(b),(c)}
    };
    \end{feynman}
  \end{tikzpicture}\right|^2}\;.
\end{equation*}
where the labels in the last two diagrams specify the ordering $t_1>t_2$. For $n=2$, each of the four terms has the form:
\begin{equation}
 \int_{t_0}^{t_{\mathrm{max}}}\mathrm{d}t_1\,\Gamma(t_1)
 \int_{t_0}^{t_1}\mathrm{d}t_2\,\Gamma(t_2)\;.
\end{equation}
The integration is over the region $t_2<t_1$. Since the integrand is symmetric under $t_1\leftrightarrow t_2$, this is half of the integral over the full square:
\begin{equation}
 \frac{1}{2!}\left(\int_{t_0}^{t_{\mathrm{max}}}\mathrm{d}t'\,\Gamma(t')\right)^2\;.
\end{equation}
More generally, for $n$ emissions, \emph{each ordered history} gives:
\begin{equation}
 \int_{t_0}^{t_{\mathrm{max}}}\mathrm{d}t_1
 \int_{t_0}^{t_1}\mathrm{d}t_2\cdots
 \int_{t_0}^{t_{n-1}}\mathrm{d}t_n\prod_{i=1}^n\Gamma(t_i)
 =\frac{1}{n!}\left(\int_{t_0}^{t_{\mathrm{max}}}\mathrm{d}t'\,\Gamma(t')\right)^n\;.
\end{equation}
There are $2^n$ assignments of the ordered emissions to the two quark lines. Therefore, within this approximation:
\begin{equation}\label{eq:omegancollinear}
 \Omega_n(t_0,t_{\mathrm{max}})
 =\frac{2^n}{n!}\left(\int_{t_0}^{t_{\mathrm{max}}}\mathrm{d}t'\,\Gamma(t')\right)^n\;.
\end{equation}
To obtain the contribution from any number of real emissions, we sum over $n$, including the Born term $\Omega_0=1$:
\begin{equation}
 \sum_{n=0}^{\infty}\sigma_n^{\mathrm{real}}(t_0)
 =\sigma_0\exp\left[2\int_{t_0}^{t_{\mathrm{max}}}\mathrm{d}t'\,\Gamma(t')\right]
 =\frac{\sigma_0}{\Delta^2(t_0,t_{\mathrm{max}})}\;,
\end{equation}
where we have defined the \textit{Sudakov form factor}:
\begin{equation}\label{eq:sudakov}
 \Delta(t_0,t_{\mathrm{max}})
 \equiv\exp\left[-\int_{t_0}^{t_{\mathrm{max}}}\mathrm{d}t'\,\Gamma(t')\right]\;.
\end{equation}

To understand its meaning, recall that the real-emission terms alone do not include the virtual and unresolved contributions. In the present example, those contributions supply the factor $\Delta^2$ that normalizes the exclusive probabilities:
\begin{equation}
 \mathcal P_n=\Delta^2(t_0,t_{\mathrm{max}})\Omega_n(t_0,t_{\mathrm{max}})\;,
 \qquad \sum_{n=0}^{\infty}\mathcal P_n=1\;.
\end{equation}
The corresponding exclusive cross sections are $\sigma_n^{\mathrm{excl}}=\sigma_0\mathcal P_n$, so their sum is $\sigma_0$, as required by shower unitarity. In particular, $\mathcal P_0=\Delta^2$ is the probability that \emph{neither} of the two outgoing quark lines emits. Thus $\Delta(t_0,t_{\mathrm{max}})$ itself is the probability of no resolved emission from \emph{one} line as it evolves from $t_{\mathrm{max}}$ down to $t_0$.

We can also obtain this interpretation directly. Let $\bar{\mathcal P}(t<t'\leq t_{\mathrm{max}})$ denote the probability of no emission from one line in this interval. If we divide the interval into $m$ pieces of width $\delta t=(t_{\mathrm{max}}-t)/m$, with $t_i=t+i\delta t$, then:
\begin{equation}
 \bar{\mathcal P}(t<t'\leq t_{\mathrm{max}})
 =\lim_{m\rightarrow\infty}\prod_{i=0}^{m-1}
 \left[1-\Gamma(t_i)\delta t+\mathcal O(\delta t^2)\right]\;.
\end{equation}
Here each factor is the probability of no emission in one small interval. Taking the logarithm, using $\ln(1-x)=-x+\mathcal O(x^2)$, and then taking the limit replaces the sum by an integral:
\begin{equation}
 \bar{\mathcal P}(t<t'\leq t_{\mathrm{max}})
 =\exp\left[-\int_t^{t_{\mathrm{max}}}\mathrm{d}t'\,\Gamma(t')\right]
 =\Delta(t,t_{\mathrm{max}})\;.
\end{equation}
Since $\Gamma$ is non-negative, $0\leq\Delta\leq1$. Moreover, splitting the interval at an intermediate scale gives:
\begin{equation}
 \Delta(t_0,t_{\mathrm{max}})
 =\Delta(t_0,t_1)\Delta(t_1,t_{\mathrm{max}})\;,
 \qquad t_0<t_1<t_{\mathrm{max}}\;.
\end{equation}
The example above, in which emissions do not change the rate, gives a Poisson distribution for their number. In a shower, however, each branching changes the daughter energies and available phase space and can introduce new radiating partons. We therefore apply the Sudakov construction to the current state after each branching; the full shower multiplicity need not be Poissonian.

We now wish to find the probability that the \emph{next} emission occurs at $t$, starting from $t_{\mathrm{max}}$. For this to happen, the line must reach $t$ without emitting and then emit in $\mathrm{d}t$. Hence:
\begin{equation}\label{eq:emissionatTstart}
 \mathrm{d}\mathcal P(\mathrm{next~emission~at}~t)
 =\Gamma(t)\mathrm{d}t\times\bar{\mathcal P}(t<t'\leq t_{\mathrm{max}})\;,
\end{equation}
and substituting the no-emission probability, we finally obtain:
\begin{equation}\label{eq:pnextemission}
 \mathrm{d}\mathcal P(\mathrm{next~emission~at}~t)
 =\mathrm{d}t\,\Gamma(t)\Delta(t,t_{\mathrm{max}})\;.
\end{equation}
There is also a finite probability of reaching $t_0$ without any emission. We can check that the two outcomes exhaust all possibilities:
\begin{equation}\label{eq:unitarity}
 \Delta(t_0,t_{\mathrm{max}})
 +\int_{t_0}^{t_{\mathrm{max}}}\mathrm{d}t\,\Gamma(t)\Delta(t,t_{\mathrm{max}})=1\;,
\end{equation}
which follows from $\partial_t\Delta(t,t_{\mathrm{max}})=\Gamma(t)\Delta(t,t_{\mathrm{max}})$. The Monte Carlo algorithm must therefore sample both the next-emission scale and the possibility of no emission above the cut-off.

\subsection{The Parton Shower Algorithm}

Equation~\ref{eq:pnextemission} gives the algorithm for a single evolving line, shown in Algorithm~\ref{alg:partonshower}. Conditional on an emission at $t$, the normalized density for $z$ is
\begin{equation}
 f(z\mid t)=\frac{\alpha_S(\mu(t,z))P(z)}{2\pi t\Gamma(t)}\;,
 \qquad z_-(t)<z<z_+(t)\;.
\end{equation}
\begin{algorithm}
\begin{algorithmic}[1]
\Procedure{Evolve one line}{$t_{\mathrm{max}},t_0$}
\Loop
\State Choose a uniform random number $R\in(0,1)$
\If{$R\leq\Delta(t_0,t_{\mathrm{max}})$}
\State \textbf{stop}: no further resolved branching
\EndIf
\State Solve $R=\Delta(t,t_{\mathrm{max}})$ for $t$
\State Sample $z$ from $f(z\mid t)$ and record the branching
\State Set the new upper scale from $t$ and the daughter kinematics
\EndLoop
\EndProcedure
\end{algorithmic}
\caption{Evolution of one line. For pure ordering, the new upper scale is $t$; for the continuing quark in the angular-ordered construction of section~\ref{sec:shower}, it is $z^2t$.}\label{alg:partonshower}
\end{algorithm}
Solving for $t$ and sampling $z$ directly can be difficult when the coupling and phase-space boundaries depend on the branching variables. The Sudakov veto algorithm provides an exact sampling method for the chosen approximate shower probability. For an extension using interleaved resampling to improve statistical convergence, see~\cite{Olsson:2019wvr}.

\subsection{The Sudakov Veto Algorithm}\label{sec:SudakovVeto}

During the parton shower, generated via Algorithm~\ref{alg:partonshower}, we obtain successive values of $t$ by solving $R = \Delta(t, t_\mathrm{max})$. To achieve this, we start by taking the natural logarithm of the equation:
\begin{equation}
    \ln R = - \int_t^{t_\mathrm{max}} \mathrm{d} t' \Gamma(t') \;.
\end{equation}
We then define:
\begin{equation}
\tilde{\Gamma}(t) = \int^t \mathrm{d} t' \Gamma(t')\;,
\end{equation}
and therefore:
\begin{equation}
\ln R = - \left[ \tilde{\Gamma}(t_\mathrm{max}) - \tilde{\Gamma}(t) \right]\;.
\end{equation}
Solving for $t$:
\begin{equation}
t = \tilde{\Gamma}^{-1} \left[ \ln R + \tilde{\Gamma}(t_\mathrm{max})\right]\;.
\end{equation}
In most cases, $\Gamma(t)$ cannot be integrated analytically. Even when it can, finding the inverse, $\tilde{\Gamma}^{-1}$, may be non-trivial. One solution is to choose a simpler function, $\hat{\Gamma}$, that has an analytic primitive with an easily computed inverse and overestimates $\Gamma$ everywhere:
\begin{equation}
\hat{\Gamma}(t) \geq \Gamma(t)~ \forall t\;.
\end{equation}
Using this function, we can construct a trial Sudakov factor:
\begin{equation}\label{eq:SudakovUnderestimate}
    \hat{\Delta}(t,t_\mathrm{max}) = \exp\left\{ - \int_t^{t_\mathrm{max}} \mathrm{d} t' \hat{\Gamma}(t') \right\}\;.
\end{equation}
Since $\hat{\Gamma}$ is chosen to be an overestimate of $\Gamma$ everywhere, $\hat{\Delta}$ is an \textit{underestimate} of $\Delta$ everywhere. Since the Sudakov represents the no-emission probability, we expect $\hat{\Delta}$ to describe more emissions than $\Delta$. Therefore, the strategy is to generate $t$ according to $\hat{\Delta}$ and `correct' by accepting the generated value of $t$ with probability $\Gamma(t)/\hat{\Gamma}(t)$. If the value is rejected, we then replace $t_\mathrm{max}$ with the rejected value of $t$, despite the fact that no emission has occurred, and continue the evolution. This is known as the `Sudakov veto algorithm' (SVA).\footnote{See also~\cite{Platzer:2011dq} for a general version of the algorithm, directly including the dependence on the infrared cut-off.} We defer the proof of this method to appendix~\ref{app:sudakovproof}.

To implement the Sudakov veto algorithm, we seek the next emission scale, $t$, as the solution of

\begin{equation}
\hat{\Delta}(t, t_\mathrm{max}) = \exp\left\{ - \int^{t_\mathrm{max}}_{t} \mathrm{d}t' \hat{\Gamma} (t') \right\} = R\;,
\end{equation}
or:
\begin{equation}
\int^{t_\mathrm{max}}_{t} \mathrm{d}t' \hat{\Gamma} (t') = \ln\left(\frac{1}{R}\right)\;.
\end{equation}
We now define a function:
\begin{equation}
    \rho({\hat{z}_+},{\hat{z}_-}) = \int_{\hat{z}_-}^{\hat{z}_+} \mathrm{d} z \frac{\hat\alpha_S}{2\pi} \hat{P}(z) = t\hat{\Gamma} (t)\;,
\end{equation}
to obtain:
\begin{equation}
    \int^{t_\mathrm{max}}_{t} \frac{\mathrm{d}t'}{t'} \rho = \ln\left(\frac{1}{R}\right)\;,
\end{equation}
or:
\begin{equation}
\int^{t_\mathrm{max}}_{t} \mathrm{d}\ln t' \rho = \ln\left(\frac{1}{R}\right)\;.
\end{equation}
If the proposal bounds in $\rho({\hat{z}_+},{\hat{z}_-})$ do not depend on $t'$ and a constant overestimate of $\alpha_S$, denoted by $\hat{\alpha}_S$, is used, then:
\begin{equation}
\left. \rho \ln t' \right|_t^{t_\mathrm{max}} = \ln\left(\frac{1}{R}\right) \Rightarrow \rho \ln \frac{t_\mathrm{max}}{t}  = \ln\left(\frac{1}{R}\right)\;,
\end{equation}
which yields after exponentiation $t =  t_\mathrm{max} R^{1/\rho}$. The same solution can also be obtained as the root of an equation in the logarithm of $t$:
\begin{equation}
\ln \frac{t}{t_\mathrm{max}} = \frac{1}{\rho} \ln R\;.
\end{equation}
We therefore define the function
\begin{equation}
E(t) = \ln \frac{t}{t_\mathrm{max}} - \frac{1}{\rho} \ln R\;,
\end{equation}
where, explicitly,
\begin{equation}
\rho({\hat{z}_+},{\hat{z}_-}) = \int_{\hat{z}_-}^{\hat{z}_+} \mathrm{d} z \frac{\hat{\alpha}_S}{2\pi} \hat{P}(z) = \int^{\hat{z}_+} \mathrm{d} z \frac{\hat{\alpha}_S}{2\pi} \hat{P}(z) - \int^{\hat{z}_-} \mathrm{d} z \frac{\hat{\alpha}_S}{2\pi} \hat{P}(z)\;.
\end{equation}
If we also define:
\begin{equation}
\tilde{\rho}(z) = \int^{z} \mathrm{d} z \frac{\hat{\alpha}_S}{2\pi} \hat{P}(z)\;,
\end{equation}
then $\rho({\hat{z}_+},{\hat{z}_-}) = \tilde{\rho}(\hat{z}_+)  - \tilde{\rho}(\hat{z}_-)$, where $\hat{z}_+ > z_+$ and $\hat{z}_- < z_-$ overestimate the integration region in $z$.
The next value of $t$ is then calculated by solving $E(t)=0$:
\begin{equation}\label{eq:nextvalue}
E(t) = \ln \frac{t}{t_\mathrm{max}} - \left[\frac{1}{\tilde{\rho}(\hat{z}_+)  - \tilde{\rho}(\hat{z}_-)}\right] \ln R = 0\;.
\end{equation}
This form assumes fixed proposal bounds during a scale draw. When the bounds depend on scale, the integral of $\hat\Gamma(t')$ must be evaluated before inversion.\footnote{The single-line tutorial script, \texttt{pyresias\_test.py}, provides a \texttt{--method Numerical} option that generates trial scales by numerically integrating and inverting the trial Sudakov factor with scale-dependent proposal bounds.}

To obtain the $z$ variable, we use the same dimensionless overestimate $\hat P(z)\geq P(z)$ and coupling bound $\hat\alpha_S\geq\alpha_S(\mu)$ as in the scale proposal. Thus
\begin{equation}
 \hat\Gamma(t)=\frac{1}{t}\int_{\hat z_-}^{\hat z_+}\mathrm{d}z\,
 \frac{\hat\alpha_S}{2\pi}\hat P(z)\;.
\end{equation}
The proposal interval is held fixed during one analytic scale draw; its excess over the physical region at the proposed scale is removed by a phase-space veto.

We then need to solve:
\begin{equation}
    \rho (z,{\hat{z}_-}) = R' \rho({\hat{z}_+},{\hat{z}_-}),
\end{equation}
for $z$, where $R'$ is another random number in $[0,1]$. We write the equation for $z$ as:
\begin{equation}
    \tilde{\rho}(z) - \tilde{\rho}({\hat{z}_-}) = R' \left[ \tilde{\rho}({\hat{z}_+}) - \tilde{\rho}({\hat{z}_-}) \right]\;,
\end{equation}
and we obtain $z$ by solving:
\begin{equation}
    z = \tilde{\rho}^{-1} \left\{ \tilde{\rho}({\hat{z}_-}) + R' \left[\tilde{\rho}({\hat{z}_+}) - \tilde{\rho}({\hat{z}_-}) \right] \right\}\;,
\end{equation}
where $\tilde{\rho}^{-1}$ is the inverse of $\tilde{\rho}$. The candidate is accepted only inside the physical phase space and with probability $[P(z)/\hat P(z)][\alpha_S(\mu)/\hat\alpha_S]$. The coupling bound must cover every allowed coupling scale. For the frozen running coupling used below, its value at the freeze scale provides such a bound. After a veto, the evolution resumes from the rejected scale; the further reduction of the upper scale described in Algorithm~\ref{alg:partonshower} applies only after an accepted branching.

Finally, the SVA can be extended to several competing branching channels, as we discuss in section~\ref{sec:competition}.

\subsection{Soft Emissions}\label{sec:soft}

So far, we have considered collinear emissions in the parton shower, including emissions that are both soft and collinear. Large-angle soft emissions are also important and can be treated through angular ordering, which incorporates colour coherence. It is instructive to study a related phenomenon in the context of electrodynamics, where angular ordering accounts for the suppression of soft radiation from $e^+ e^-$ pairs (the Chudakov effect). We examine a simple heuristic explanation inspired by Ref.~\cite{Ellis:1996mzs}.

\begin{figure}[!htp]
\centering

\includegraphics[]{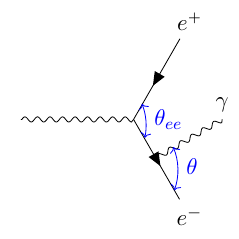}
\caption{Photon emission from an $e^+e^-$ pair with opening angle $\theta_{ee}$. Both angles are taken to be small: $\theta, \theta_{ee} \ll 1$.}
\label{fig:eephoton}
\end{figure}

Consider the emission of a soft photon at an angle $\theta$ from an electron in an $e^+ e^-$ pair with opening angle $\theta_{ee}$ (see fig.~\ref{fig:eephoton}). Both angles are taken to be small: $\theta, \theta_{ee} \ll 1$.
The four-momenta can be written as
\begin{eqnarray}
    e^-~\text{before emission:}~p_{e^*} &=& (E_{e^*}, p, \mathbf{0})\;, \\\nonumber
    \text{emitted photon:}~p_\gamma &=& (E_\gamma , zp, \mathbf{k}_T)\;, \\\nonumber
    e^-~\text{after emission:}~p_{e} &=& (E_{e}, (1-z)p, -\mathbf{k}_T)\;,
\end{eqnarray}
where $z$ is the momentum fraction of the emitted photon and $\mathbf{k}_T$ its transverse momentum. Neglecting the electron mass, the squared energies are
\begin{eqnarray}
E_\gamma^2 &=& (zp)^2 + \mathbf{k}_T^2\;, \\\nonumber
E_{e*}^2 &=& p^2\;, \\\nonumber
E_e^2 &=& (1-z)^2 p^2 + \mathbf{k}_T^2\;.
\end{eqnarray}
The energy denominator for this intermediate electron--photon state, relative to the on-shell parent energy $p$, is then:
\begin{eqnarray}
    \Delta E &=& E_e + E_\gamma - E_{e*} \\ \nonumber
    &=& \sqrt{(zp)^2 + \mathbf{k}_T^2} + \sqrt{ (1-z)^2 p^2  + \mathbf{k}_T^2 } - p\;,
\end{eqnarray}
which, after expansion for small transverse momentum, $|\mathbf{k}_T| \ll zp, (1-z)p$, becomes:
\begin{equation}
    \Delta E \approx zp \left( 1 + \frac{1}{2} \frac{\mathbf{k}_T^2}{z^2p^2}\right) + (1-z) p\left(1+ \frac{1}{2} \frac{\mathbf{k}_T^2}{p^2(1-z)^2}\right) - p\;,
\end{equation}
which leads to:
\begin{equation}
    \Delta E \approx \frac{1}{2} \frac{\mathbf{k}_T^2 }{p} \frac{1}{z(1-z)}\;.
\end{equation}
For soft emissions, $z(1-z) \sim z$, so:
\begin{equation}
    \Delta E \sim \frac{\mathbf{k}_T^2}{zp}\;.
\end{equation}
By considering the right-angled triangle formed by the outgoing electron and the emitted photon ($\sim zp$ at an angle $\theta$), we also have $|\mathbf{k}_T| \approx zp \sin \theta \approx zp \theta$, which gives
\begin{equation}
\Delta E \sim z p \theta^2.
\end{equation}
Due to the uncertainty principle, the time available for photon emission is $\Delta t \sim 1/\Delta E$. During this time, the transverse separation of the $e^+ e^-$ pair becomes $\Delta b \sim \theta_{ee} \Delta t$. For the photon-emission probability to be non-negligible, the photon must be able to resolve this separation. This will be the case if $\Delta b > \lambda / \theta$, where $\lambda$ is the photon wavelength and $\lambda \sim 1/E_\gamma \sim 1/(zp)$. This leads to the condition $\Delta b > 1/(zp\theta)$. Substituting for $\Delta b$, we obtain:
\begin{equation}
\frac{\theta_{ee}}{zp \theta^2} > \frac{1}{zp \theta} \Rightarrow \theta_{ee} > \theta\;.
\end{equation}
Photons emitted at angles $\theta > \theta_{ee}$ cannot resolve the individual $e^+$ and $e^-$ charges, i.e.\ they only see the net zero charge of the system. This implies that soft photon emission is suppressed at angles $\theta > \theta_{ee}$. In QCD, a soft gluon at an angle larger than the opening angle of a collinear pair probes the pair's total colour charge, which is the parent's charge and need not vanish. The corresponding interference can be organized into angular-ordered branching probabilities after the appropriate azimuthal average~\cite{Marchesini:1983bm}. Thus coherence does not generally mean an absence of wide-angle radiation: that radiation belongs to the unresolved parent system.

\subsection{Connections to Shower Development}\label{sec:scope}

The ingredients developed thus far connect directly to questions that arise in precision shower calculations. In particular, an ordering variable specifies which branching occurs first, while a recoil prescription determines how later branchings modify momenta already generated. Their combined effect matters for logarithmic accuracy. Studies of angular-ordered reconstruction~\cite{Bewick:2019rbu} and the development of showers with demonstrated next-to-leading-logarithmic accuracy~\cite{Dasgupta:2020fwr,Herren:2022jej}, and more recently beyond it~\cite{vanBeekveld:2024wws}, provide concrete examples. Reproducing a splitting kernel or using a two-loop coupling alone does not establish that accuracy for an observable.

Angular-ordered and dipole constructions organize coherent radiation differently. In the implementation below, branchings are generated first and global energy--momentum conservation is restored during reconstruction. In a Catani--Seymour shower, an emitter and spectator define a local momentum map~\cite{Schumann:2007mg}; other showers use different recoil prescriptions. The present choice is useful for separating branching probabilities from reconstruction in a tutorial. A comparison between implementations is meaningful only after their kernels, phase space, coupling, and recoil choices have been specified.

The singular approximations do not describe arbitrary hard, widely separated partons accurately. Matching and merging combine showers with fixed-order matrix elements while controlling double counting. Original examples include matching to next-to-leading-order calculations~\cite{Frixione:2002ik,Nason:2004rx} and merging different tree-level parton multiplicities~\cite{Catani:2001cc}. Neither is implemented here. The available source code instead permits exercises in changing a kernel, a cut-off, or a recoil choice and identifying which observable responds.

We also omit initial-state radiation, parton distributions, hadronization, and multiple parton interactions. Backward evolution introduces the parton-distribution dependence needed for incoming hadrons; non-perturbative models describe physics below the shower cut-off. The generator reviews and manuals~\cite{Buckley:2011ms,Campbell:2022qmc,Bierlich:2022pfr,Sherpa:2024mfk,Bellm:2025pcw} provide entry points to these subjects and to spin and subleading-colour effects. The veto algorithm itself also underlies the estimation of shower uncertainties: since every accepted and rejected trial has a known probability, events can be reweighted exactly to a modified kernel or coupling without being generated again~\cite{Bellm:2016voq,Mrenna:2016sih}. Efficient sampling is another research direction; the GPU implementation of Ref.~\cite{Seymour:2024fmq} illustrates how shower evolution can be organized for parallel hardware.

\section{A Parton Shower}\label{sec:shower}

\subsection{Implementation Specifics}

The first stage of \texttt{Pyresias} simulates multiple gluon emissions from an outgoing quark (or antiquark) line:
\begin{equation*}
  \begin{tikzpicture}[baseline=-\the\dimexpr\fontdimen22\textfont2\relax]
    \begin{feynman}[inline=(a)]
    \vertex (a) {};
    \vertex [right=of a] (b);
    \vertex [above right=of b] (d);
    \vertex [above right=of c] (e);
    \vertex [right=of b] (c);
    \vertex [above right=of c] (e);
    \vertex [right=of c] (f);
    \vertex [above right=of f] (g);
    \vertex [right=of f] (h);
    \diagram*[large] {
       (a) -- [fermion] (b) -- [fermion] (c) -- [fermion] (f) -- [fermion] (h),
       (b) -- [gluon] (d),
       (c) -- [gluon] (e),
       (f) -- [gluon] (g),
    };
    \node[align=left] at (2.9,0.4) {...};
    \node[align=left] at (1.9,0.4) {...};
    \end{feynman}
  \end{tikzpicture}
\end{equation*}
This treatment includes the collinear enhancements and treats soft wide-angle emissions through angular ordering (i.e.\ coherent branching). We perform the branchings in a frame in which the parent quark moves along the $z$ direction and then rotate (see appendix~\ref{app:rotations}) all the reconstructed momenta into the original event frame. The supplied event implementations assume a massless $e^+e^-\rightarrow q\bar q$ hard event in its centre-of-mass frame.

For coherent branching, the corresponding evolution variable is~\cite{Ellis:1996mzs}: $\tilde{t} = 2E^2 \zeta$, for a parton of energy $E$, where $\zeta = \frac{p_b . p_c}{E_b E_c} \simeq 1 - \cos \theta$ in a branching $a \rightarrow bc$. The factor of two fixes our convention so that $\tilde t\simeq E^2\theta^2$, consistent with the mass and transverse-momentum relations below. For this variable, the angular-ordering condition becomes

\begin{equation}
\tilde{t}_b < z^2 \tilde{t}\;,\qquad \tilde{t}_c < (1-z)^2 \tilde{t}\;.
\end{equation}
Thus, for example, in $q \rightarrow qg$ the outgoing quark and gluon start their evolution in $\tilde{t}$ at the respective upper bounds above. In the introductory sampler, we choose the symmetric proposal limits:
\begin{equation}
    z_+ = 1 - \sqrt{ \frac{\tilde{t}_0}{\tilde{t}}}\;;~ z_- =  \sqrt{ \frac{\tilde{t}_0}{\tilde{t}}}\;.
\end{equation}
Here $\tilde t_0=Q_c^2$ fixes the proposal cut-off. These bounds are supplemented by the physical condition $p_T\geq Q_c$; they should not be interpreted as a consequence of angular ordering alone. For a branching with massless on-shell daughters, the collinear kinematic relations are:
\begin{itemize}
    \item the virtual mass squared of $a$ in $a\rightarrow bc$ is given by $m_a^2 = z(1-z) \tilde{t}$;
    \item the transverse momentum squared of the branching is given by $p_T^2 = z^2 ( 1-z)^2 \tilde{t}$.
\end{itemize}
From here on, as in the code, we write $t$ for the evolution variable $\tilde t$ and $t_0$ for $\tilde t_0$. In the running-coupling option, the strong coupling is evaluated at the transverse momentum of the branching, $\mu=\max(p_T,Q_c)$, as this captures some of the next-to-leading-order behaviour of the branching probability~\cite{Amati:1980ch}. This choice alone does not make the shower accurate at next-to-leading-logarithmic order. The notebook also offers a fixed-coupling example. The values of $\alpha_S$ are calculated at next-to-leading order following Refs.~\cite{hocheprestel,Seymour:2024fmq}. Finally, the overestimate for the $q \rightarrow qg$ splitting function is given by:
\begin{equation}\label{eq:qgqover}
\hat{P}(z) = C_F \frac{2}{1-z}\;,
\end{equation}
which satisfies $\hat{P}(z) \geq P(z)\quad(0<z<1)$. For running coupling, the bound is its frozen value at the cut-off; for fixed coupling, the bound equals the chosen constant. The above overestimate of eq.~\ref{eq:qgqover} implies that:
\begin{equation}
\tilde{\rho}(z) = -\frac{\hat\alpha_S C_F}{\pi}\ln(1-z)\;.
\end{equation}
The factor $\tilde{\rho}(\hat{z}_+) - \tilde{\rho}(\hat{z}_-)$ in the denominator of eq.~\ref{eq:nextvalue} must be positive for a solution for $t$ to exist. This condition becomes $ - \log( \sqrt{t_0/t} ) + \log(1-\sqrt{t_0/t}) > 0 $, which in turn implies that $( 1-\sqrt{t_0/t} )/\sqrt{t_0/t} > 1 $, and finally that $t > 4 t_0$. This is the threshold for a non-empty proposal interval. The additional condition $p_T=z(1-z)\sqrt{t}\geq Q_c$ requires $t\geq16Q_c^2$ for a physical branching, since $z(1-z)\leq1/4$. Trials in the excess region are vetoed.

The proposal can be implemented without a numerical inverse. At a current upper scale $T$, choose fixed bounds $\hat z_-=Q_c/\sqrt T$ and $\hat z_+=1-\hat z_-$, with $\hat z_-<\hat z_+$, and define
\begin{equation}
 \rho=\frac{\hat\alpha_SC_F}{\pi}\ln\frac{1-\hat z_-}{1-\hat z_+}\;.
\end{equation}
Two independent uniform random numbers $R_1,R_2\in(0,1)$ give
\begin{equation}\label{eq:workedproposal}
 t=T R_1^{1/\rho}\;,
 \qquad z=1-(1-\hat z_-)\left(\frac{1-\hat z_+}{1-\hat z_-}\right)^{R_2}\;.
\end{equation}
Check the physical bounds and transverse-momentum cut-off at this new $t$. If they are satisfied, accept the candidate with probability
\begin{equation}
 \frac{1+z^2}{2}\,\frac{\alpha_S(\mu)}{\hat\alpha_S}\;.
\end{equation}
This can be implemented as two independent random-number tests. A veto sets $T=t$ and repeats the proposal. An accepted quark emission is recorded, and the continuing quark starts at $T=z^2t$. This distinction between a veto and an accepted branching is essential. The functions \texttt{Get\_tEmission\_direct}, \texttt{Get\_zEmission}, \texttt{Generate\_Emission}, and \texttt{Evolve} in \texttt{pyresias\_test.py} implement these steps; the notebook walks through the same construction.

\subsection{Global Energy--Momentum Conservation}\label{sec:globcons}
We wish to preserve the total energy of the system in the centre-of-mass frame of the hard collision. If the momenta of the parent partons before evolution are $p_J = (\mathbf{p}_J, \sqrt{\mathbf{p}_J^2 + m_J^2})$ in this frame, then they satisfy:
\begin{equation}
    \sum_{j=1}^n \sqrt{\mathbf{p}_J^2 + m_J^2} = \sqrt{s}\;,
\end{equation}
and the sum of the spatial momenta is zero: $\sum_{j=1}^n \mathbf{p}_J = \mathbf{0}$. The jet parents have momenta $q_J = \left(\mathbf{q}_J, \sqrt{\mathbf{q}_J^2 + q_J^2}\right)$ after the parton shower. We need to restore momentum conservation in a way that disturbs the internal structure of the jets as little as possible. This is achieved by first rotating the outgoing jet into the original jet direction and then boosting it along its axis such that the momenta are rescaled by a common factor $k$, determined by:
\begin{equation}
    \sum_{j=1}^n \sqrt{k^2 \mathbf{p}_J^2 + q_J^2} = \sqrt{s}\;.
\end{equation}
This equation can be solved analytically for two jets or, in general, numerically, e.g.\ via Newton--Raphson root finding. The rotations and boosts need to be applied to each jet in the centre-of-mass frame of the collision. Schematically:
\begin{equation}
    q_J = \left(\mathbf{q}_J, \sqrt{\mathbf{q}_J^2 + q_J^2}\right) \xrightarrow{rot. + boost} q_J' = \left( k \mathbf{p}_J, \sqrt{ k^2 \mathbf{p}_J^2 + q_J^2}\right)\;.
\end{equation}
See appendix~\ref{app:rotations} for details on the rotations. The \texttt{Python} code \texttt{pyresias.py} implements this reconstruction approach for the case of $e^+ e^- \rightarrow q\bar{q}$. Within each jet, it assigns the momenta in a simplified way: each emitted gluon receives the sampled transverse momentum at a random azimuth, while the continuing quark stays on the jet axis, with its longitudinal momentum reduced by a factor $z$ at each emission. The sampled $p_T$ and virtual mass are therefore not exactly the relative transverse momentum and invariant mass of the reconstructed partons, although the global rescaling conserves the total four-momentum. The $\tilde q$ shower described next replaces this assignment by a reconstruction in the Sudakov basis that preserves the transverse momenta of the branchings.

\subsection{The \texorpdfstring{$\tilde{q}$}{q-tilde} Shower}\label{sec:qtilde}

We now describe an alternative implementation, following the angular-ordered shower of Ref.~\cite{Gieseke:2003rz}, as implemented in \texttt{Herwig}~\cite{Bahr:2008pv,Bellm:2025pcw}. The corresponding code is provided in \texttt{pyresias\_qtilde.py}. As before, we restrict ourselves to gluon radiation from massless quark and antiquark lines.

\subsubsection{The \texorpdfstring{$\tilde{q}$}{q-tilde} Evolution Variable}

For a branching $a\rightarrow bc$, where $b$ is the outgoing quark and $c$ is the gluon, we define:
\begin{equation}\label{eq:qtilde}
    \tilde{q}^2 = \frac{p_T^2}{z^2(1-z)^2}\;,
\end{equation}
where $z$ is the light-cone momentum fraction retained by the quark and $p_T$ is the relative transverse momentum of the branching. For a single emission with massless, on-shell daughters, the parent virtuality is $q_a^2 = z(1-z)\tilde{q}^2$. In the collinear limit, $\tilde{q}\simeq E\theta$, where $E$ is the parent energy and $\theta$ is the opening angle of the branching. Thus, angular ordering requires:
\begin{equation}
    \tilde{q}_b < z\tilde{q}_a\;,\qquad \tilde{q}_c < (1-z)\tilde{q}_a\;.
\end{equation}
Only the quark is evolved further in this first version; gluon branchings are included in the extension described in section~\ref{sec:competition}. For the process $e^+e^-\rightarrow q\bar{q}$, the initial evolution scale is set by the invariant mass of the quark and its colour partner, $\tilde{q}_{\mathrm{max}}^2=s$. We then apply the Sudakov veto algorithm of section~\ref{sec:SudakovVeto} with $t=\tilde{q}^2$ and the splitting-function overestimate of eq.~\ref{eq:qgqover}.

Emissions are accepted only if $p_T\geq p_{T,\mathrm{min}}$. The coupling is evaluated at the scale $p_T=z(1-z)\tilde{q}$ and frozen below a separate scale $Q_c$. The coupling scale and its overestimate are therefore:
\begin{equation}
    \mu=\max(p_T,Q_c)\;,\qquad \hat{\alpha}_S=\alpha_S(Q_c)\;.
\end{equation}
Using the same frozen coupling in the veto probability ensures that $\alpha_S(\mu)/\hat{\alpha}_S\leq 1$. For the comparison with \texttt{Herwig 7}, we use $p_{T,\mathrm{min}}=0.900$~GeV and $Q_c=0.935$~GeV. The two-loop coupling is fixed by $\alpha_S(M_Z)=0.1074$, with $M_Z=91.1876$~GeV, and is matched continuously at the charm, bottom, and top thresholds of $1.6$, $5.0$, and $172.69$~GeV, respectively. These thresholds enter the running of the coupling; the shower kinematics remain massless.

\subsubsection{Shower Kinematics in the Sudakov Basis}
Each showering particle in the hard subprocess is considered to be a \textit{shower progenitor}. To reconstruct the momenta from $\tilde{q}, z$, and the azimuthal angle $\phi$, chosen uniformly in $[-\pi,\pi]$, we use the Sudakov basis~\cite{Bahr:2008pv}. We take $p$ to be the original momentum of the progenitor and $n$ to be the momentum of its colour partner. For the massless $q\bar{q}$ system, $p^2=n^2=0$ and $p\cdot n>0$. In their centre-of-mass frame, we choose $p$ along the $z$ direction and $n$ in the opposite direction. Each momentum in the shower is then written as:
\begin{equation}
    q_i = \alpha_i p + \beta_i n + q_{\perp i}\;,
\end{equation}
where $p\cdot q_{\perp i}=n\cdot q_{\perp i}=0$ and $q_{\perp i}^2=-\mathbf{q}_{\perp i}^2$. The on-shell condition fixes the remaining coefficient for each final-state quark or gluon:
\begin{equation}
    \beta_i = \frac{\mathbf{q}_{\perp i}^2}{2\alpha_i p\cdot n}\;.
\end{equation}

Starting from $\alpha_a=1$ and $\mathbf{q}_{\perp a}=\mathbf{0}$, the coefficients of the daughters in each branching are obtained from:
\begin{equation}
    \alpha_b=z\alpha_a\;,\qquad \alpha_c=(1-z)\alpha_a\;,
\end{equation}
and
\begin{eqnarray}
    \mathbf{q}_{\perp b} &=& z\mathbf{q}_{\perp a}+\mathbf{k}_{\perp}\;,\nonumber\\
    \mathbf{q}_{\perp c} &=& (1-z)\mathbf{q}_{\perp a}-\mathbf{k}_{\perp}\;,
\end{eqnarray}
where $\mathbf{k}_{\perp}=p_T(\cos\phi,\sin\phi)$ is a two-dimensional vector. We repeat these steps along the quark line, retaining each emitted gluon and the final quark. The generated transverse momenta of the branchings are kept fixed when further emissions are added, while the virtualities of the intermediate quarks follow from the sums of their daughter momenta. This is the transverse-momentum-preserving reconstruction choice within the angular-ordered shower~\cite{Bewick:2019rbu}. Finally, the momenta are rotated into the original event frame, and global energy--momentum conservation is imposed as described in section~\ref{sec:globcons}.

\subsection{Competing Branching Channels}\label{sec:competition}

We now extend the shower to include $g\rightarrow gg$ and $g\rightarrow q\bar{q}$, in addition to $q\rightarrow qg$. We retain massless kinematics and allow five quark flavours in gluon splittings. The code is provided in \texttt{pyresias\_qtilde\_full.py}. The real kernels and convenient overestimates are
\begin{equation}\label{eq:allkernels}
\begin{array}{c|c|c}
\text{channel}&P_i(z)&\hat P_i(z)\\ \hline
q\rightarrow qg&C_F(1+z^2)/(1-z)&2C_F/(1-z)\\
g\rightarrow gg&C_A[1-z(1-z)]^2/[z(1-z)]&C_A/[z(1-z)]\\
g\rightarrow q\bar q&T_R[z^2+(1-z)^2]&T_R
\end{array}
\end{equation}
with $C_F=4/3$, $C_A=3$, and $T_R=1/2$. The last row is for one flavour. In the gluon-pair channel, $z$ labels either daughter, and the full symmetric interval is sampled. The kernel therefore includes a factor $1/2$ for identical daughters relative to the inclusive daughter-counting kernel $2C_A[z/(1-z)+(1-z)/z+z(1-z)]$. Each branching creates both daughters once. This convention avoids doubling the branching rate. For $g\rightarrow q\bar q$, $z$ labels the quark. The primitives of the three overestimates are $-2C_F\ln(1-z)$, $C_A\ln[z/(1-z)]$, and $T_Rz$, respectively.

A gluon can then branch through several channels, each with its own splitting probability. Writing $\Gamma_i(t)$ for the rate integrated over $z$ in channel $i$, with $t=\tilde{q}^2$, the probability of no branching in any channel is:
\begin{equation}
    \Delta_{\mathrm{tot}}(t,t_{\mathrm{max}}) = \prod_i \Delta_i(t,t_{\mathrm{max}}) = \exp\left[-\int_t^{t_{\mathrm{max}}}\mathrm{d}t'\sum_i\Gamma_i(t')\right]\;.
\end{equation}
Here, each allowed quark flavour contributes a separate $g\rightarrow q\bar{q}$ channel.

To generate the next branching, we apply the Sudakov veto algorithm independently to each available channel, starting from the same upper scale. We retain the accepted candidate with the largest value of $t$, together with its branching channel and momentum fraction $z$~\cite{Platzer:2011dq}. This gives the first branching in the downward evolution and reproduces the combined Sudakov factor above. If no channel produces a candidate above the cut-off, the parton does not branch further. After a branching at $\tilde{q}$, both daughters are evolved, with upper scales $z\tilde{q}$ and $(1-z)\tilde{q}$, respectively. The same procedure is repeated for every daughter, including secondary quarks and gluons, until their evolution ends at the cut-off. A gluon's $g\rightarrow q\bar q$ branching competes with its $g\rightarrow gg$ branching from the same angular-ordered upper scale. Since the $g\rightarrow q\bar q$ kernel has no soft singularity, soft coherence does not by itself require this restriction on the branching; we impose it, as do the default \texttt{Herwig} settings. The quark and antiquark produced in the splitting then radiate with angular-ordered upper scales, like any other daughters.

\subsection{From Branching Histories to Events}\label{sec:implementation}

The program separates probability generation from momentum reconstruction. A quark-line history is a sequence of $(t,z,\phi)$ values. With gluon branchings included, the history is a tree: each internal node stores a channel, flavour, and branching variables, and each terminal node is a final-state parton. Algorithm~\ref{alg:event} summarizes the full event construction. In the code, \texttt{Next\_Emission} completes one channel's veto loop, \texttt{Choose\_Emission} selects the highest candidate, and \texttt{EvolveParticle} builds the tree. Reconstruction then propagates the coefficients of section~\ref{sec:qtilde} to the leaves through \texttt{reconstructSudakov}.

\begin{algorithm}
\begin{algorithmic}[1]
\Procedure{Shower an event}{$p_q,p_{\bar q}$}
\State Initialize the two progenitors and their colour-partner scales
\State Put both progenitors on a list of partons awaiting evolution
\While{the list is non-empty}
\State Remove one parton and generate a candidate for each allowed channel
\If{no candidate is above the cut-off}
\State Mark this parton as a terminal node
\Else
\State Keep the candidate with the largest scale
\State Record its channel, flavour, $t$, $z$, and a uniform azimuth $\phi$
\State Create both daughters and add them with upper scales $z^2t$, $(1-z)^2t$
\EndIf
\EndWhile
\State Reconstruct the terminal momenta in the Sudakov basis
\State Rotate and boost the jets to restore the hard event's four-momentum
\State If reconstruction is impossible, resample the shower of this hard event
\State Write the final-state partons with their colour connections and event weight
\EndProcedure
\end{algorithmic}
\caption{Event construction with competing channels. In the quark-line version, emitted gluons are terminal nodes from the outset.}\label{alg:event}
\end{algorithm}

\noindent\begin{minipage}{\linewidth}
The following commands, run from the code repository after installing
its tutorial dependencies, generate and compare the two
angular-ordered examples:
\begin{lstlisting}[style=terminal]
python -m pip install -r requirements-tutorials.txt
python pyresias_qtilde.py data/eejj_ECM206.lhe.gz  -n 1000 --seed 12345 -o output/quark-lines.lhe
python pyresias_qtilde_full.py data/eejj_ECM206.lhe.gz  -n 1000 --seed 12345 -o output/all-channels.lhe
python analysis/lhe_analyzer.py output/quark-lines.lhe output/all-channels.lhe --labels QuarkLines AllChannels -o plots/channels
\end{lstlisting}
\end{minipage}
\medskip

The supplied input contains $10^4$ hard events at $206$~GeV, and a separate $10^6$-event input is used for the comparison below; \texttt{-n} counts hard events. The notebook and \texttt{pyresias.py} retain the introductory examples, and the repository README and \texttt{docs/full-shower.md} give the environment, the \texttt{Herwig} input card, and the analysis commands.

\subsection{Validation}\label{sec:results}

We first test the veto algorithm itself. For a single line, eqs.~\ref{eq:pnextemission} and~\ref{eq:unitarity} predict the distribution of the first emission scale and the probability of reaching the cut-off without an emission. The phase-space, splitting-function, and coupling vetoes must together reproduce both, whatever the overestimates. We evolve $5\times10^6$ quark lines with \texttt{Evolve} in \texttt{pyresias\_test.py}, starting from $t_{\mathrm{max}}=(206~\mathrm{GeV})^2$ with $Q_c=1$~GeV and the running coupling, and record the first accepted emission of each line. Here $t$ denotes the angular evolution variable $\tilde t$ used by this sampler.

To construct the numerical reference, we first integrate over $z$ in eq.~\ref{eq:rate}. The physical condition $z(1-z)\sqrt{t}\geq Q_c$ gives the limits:
\begin{equation}\label{eq:validation-zlimits}
 z_\pm^{\mathrm{phys}}(t)=\frac{1}{2}\left(1\pm\sqrt{1-\frac{4Q_c}{\sqrt{t}}}\right)\;,
 \qquad t\geq16Q_c^2\;,
\end{equation}
and $\Gamma(t)=0$ below this threshold. We use the quark splitting function $P(z)$ and the same coupling $\alpha_S(\max(p_T,Q_c))$ as in the sampler.\footnote{For this single-line test, the two-loop coupling is fixed by $\alpha_S(M_Z)=0.118$, with $M_Z=91.1876$~GeV, and matched at the charm and bottom thresholds of $1.25$ and $4.2$~GeV, respectively.} The $z$ integral gives $t\Gamma(t)$ directly, without using the overestimates. We then change variables to $u=\ln(t'/Q_c^2)$ in the Sudakov exponent:
\begin{equation}\label{eq:numerical-sudakov}
 \Delta(t,t_{\mathrm{max}})=\exp\left[-\int_{\ln(t/Q_c^2)}^{\ln(t_{\mathrm{max}}/Q_c^2)}
 \mathrm{d}u\,Q_c^2e^u\,\Gamma(Q_c^2e^u)\right]\;.
\end{equation}
The Jacobian cancels the explicit factor $1/t'$ in $\Gamma(t')$. Both integrals are evaluated by adaptive quadrature.\footnote{We use \href{https://docs.scipy.org/doc/scipy/reference/generated/scipy.integrate.quad.html}{\texttt{scipy.integrate.quad}}, with requested relative tolerance $10^{-10}$, the default absolute tolerance $1.49\times10^{-8}$, and a maximum of 200 subintervals for each integration.}

\begin{figure}[tbp]
\centering
\includegraphics[width=0.6\linewidth]{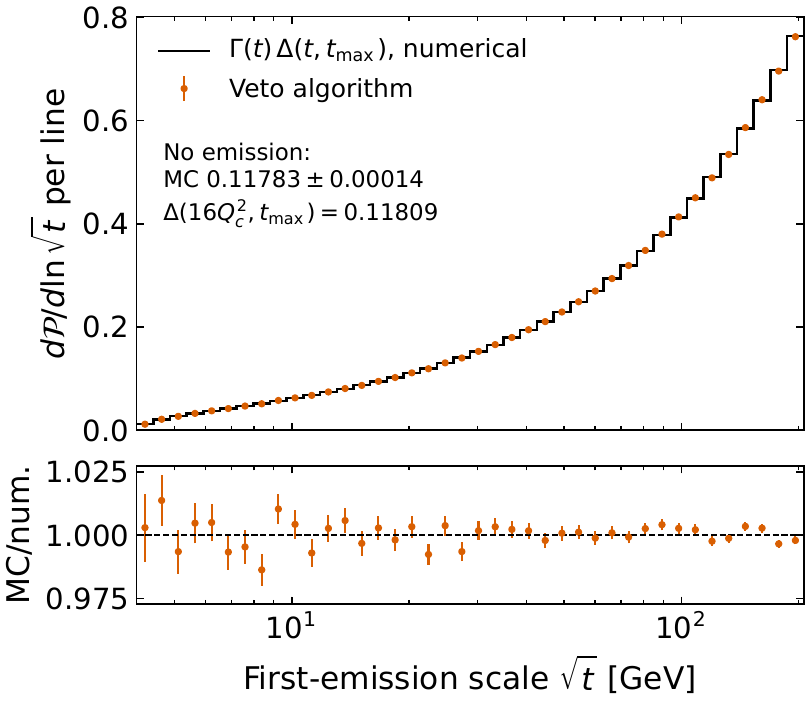}
\caption{Closure test of the veto algorithm for single quark lines, with $t_{\mathrm{max}}=(206~\mathrm{GeV})^2$, $Q_c=1$~GeV, and the running coupling. Points show the first-emission distribution for $5\times10^6$ lines evolved with \texttt{pyresias\_test.py}; the line shows the numerical bin probabilities of eq.~\ref{eq:sudakov-bin-probability}, divided by their logarithmic widths to give $\mathrm{d}\mathcal P/\mathrm{d}\ln\sqrt{t}$ per input line. The text in the upper panel compares the fraction of lines without an emission with $\Delta(16Q_c^2,t_{\mathrm{max}})$.}\label{fig:sudakov}
\end{figure}

For fig.~\ref{fig:sudakov}, we use 40 bins spaced uniformly in $\ln\sqrt{t}$ between $\sqrt{t}=4Q_c$ and $206$~GeV. We accumulate the Sudakov exponent from the upper scale downwards to obtain $\Delta$ at each bin edge. The probability in a bin with edges $t_i$ and $t_{i+1}$ is then:
\begin{equation}\label{eq:sudakov-bin-probability}
 p_i=\int_{t_i}^{t_{i+1}}\mathrm{d}t\,\Gamma(t)\Delta(t,t_{\mathrm{max}})
 =\Delta(t_{i+1},t_{\mathrm{max}})-\Delta(t_i,t_{\mathrm{max}})\;.
\end{equation}
To obtain the second equality, we differentiate the Sudakov factor with respect
to its lower scale, giving
$\partial_t\Delta(t,t_{\mathrm{max}})=\Gamma(t)\Delta(t,t_{\mathrm{max}})$.
The positive sign follows because differentiating the lower integration limit
cancels the minus sign in the exponent. Integrating this derivative over the
bin gives the difference of endpoint values.

We divide $p_i$ by the logarithmic bin width $\ln(\sqrt{t_{i+1}}/\sqrt{t_i})$ to obtain the plotted density. The Monte Carlo counts are divided by the total number of input lines and by the same bin widths. The remaining probability is that of no emission, $p_{\mathrm{none}}=\Delta(16Q_c^2,t_{\mathrm{max}})$, so that $p_{\mathrm{none}}+\sum_i p_i=1$.

Figure~\ref{fig:sudakov} shows the result. The fraction of lines without an emission is $0.11783\pm0.00014$, compared with $\Delta(16Q_c^2,t_{\mathrm{max}})=0.11809$, and the first-emission distribution and the no-emission fraction together give $\chi^2=51.1$ for 40 degrees of freedom ($p=0.11$). Here the 40 emission bins and the no-emission category have one overall normalization constraint.

Automated tests of the full shower also make the corresponding comparison for competing channels at fixed coupling, including the channel fractions. They also check the overestimate bounds, daughter ordering, and conservation of four-momentum, colour, and net flavour. Switching off gluon branchings in the full implementation reproduces the quark-line implementation event by event for the same seed.

We then compare reconstructed events with the angular-ordered shower
of \texttt{Herwig 7.3.0}~\cite{Bahr:2008pv,Bewick:2023tfi}. Both
programs shower the same $10^6$ parton-level $e^+e^-\rightarrow q\bar
q$ events, generated with \texttt{MadGraph5\_aMC@NLO}
2.9.19~\cite{Alwall:2014hca} at $\sqrt{s}=206$~GeV without primary
bottom quarks and passed to both programs in the Les Houches Event
format~\cite{Alwall:2006yp}. The comparison includes $q\rightarrow qg$, $g\rightarrow gg$, and $g\rightarrow q\bar q$ for five massless secondary flavours. Both showers use the coupling, cut-offs, and transverse-momentum-preserving reconstruction of section~\ref{sec:qtilde}.

\begin{figure}[tbp]
\centering
\includegraphics[width=0.48\linewidth]{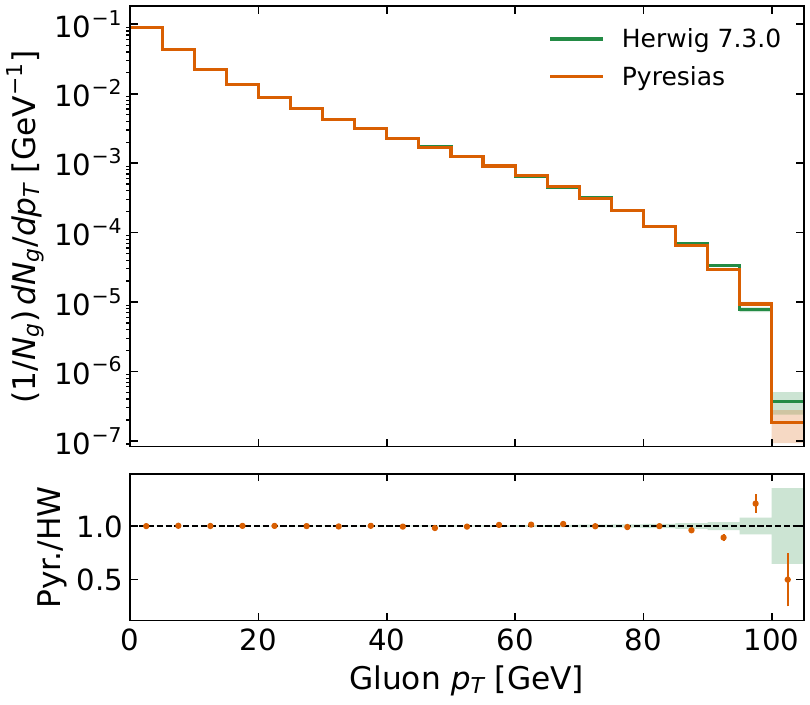}
\includegraphics[width=0.48\linewidth]{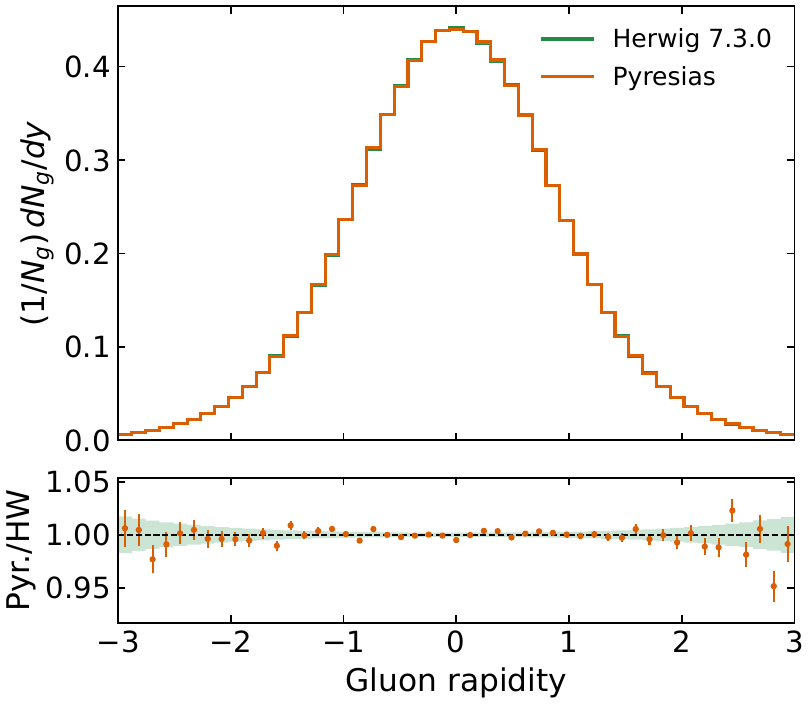}
\caption{Gluon transverse momentum (left) and rapidity (right) with respect to the beam axis, from \texttt{Herwig 7.3.0} and \texttt{Pyresias} at $206$~GeV, with $10^6$ events per generator and all three QCD branching channels. Spectra are normalized to the total number of gluons and divided by the bin width. Bands and error bars show statistical uncertainties; in the lower panels, both samples are divided by the \texttt{Herwig} central value.}\label{fig:hwvspyr}
\end{figure}

In the matching \texttt{Herwig} configuration, all three channels are angular ordered, as in the default settings, while the transverse-momentum-preserving scheme replaces the default dot-product-preserving one~\cite{Bewick:2019rbu} and the additional final-final reconstruction weight is disabled. Hadronization, photon and initial-state radiation, matrix-element corrections, and spin and soft azimuthal correlations are also switched off. This is therefore a comparison with a deliberately restricted configuration, not with a phenomenological tune. Figure~\ref{fig:hwvspyr} shows the normalized gluon transverse-momentum and rapidity distributions, measured with respect to the beam axis.

\begin{figure}[tbp]
\centering
\includegraphics[width=0.48\linewidth]{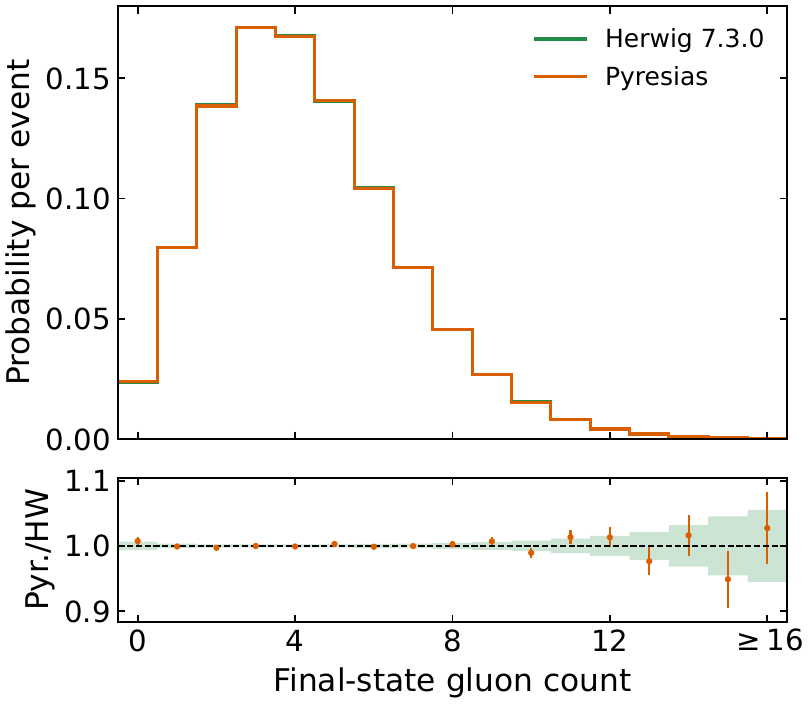}
\includegraphics[width=0.48\linewidth]{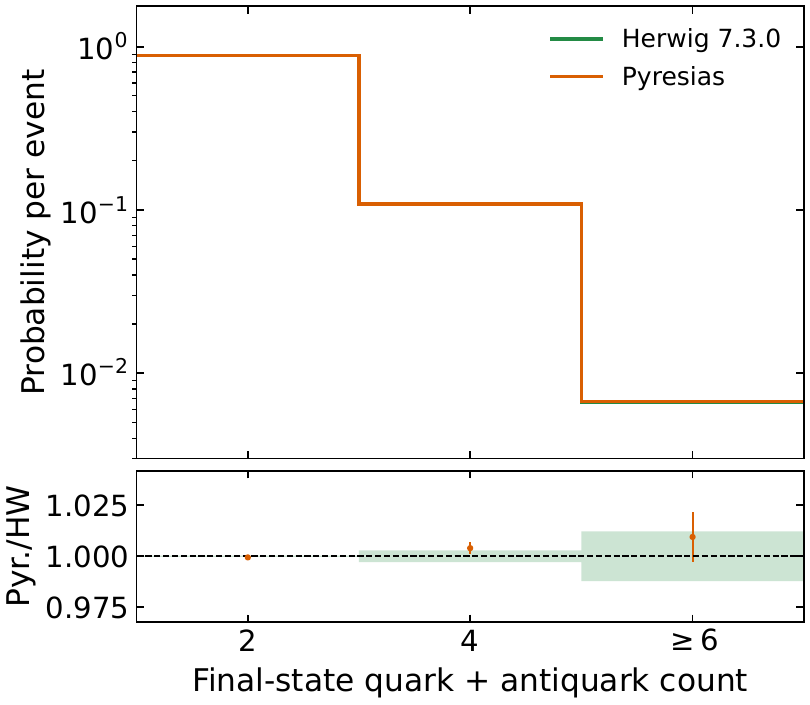}
\caption{Gluon (left) and quark-plus-antiquark (right) multiplicities for the samples of fig.~\ref{fig:hwvspyr}. The last category contains all events with at least 16 gluons (left) or at least six quarks and antiquarks (right). The lower panels are as in fig.~\ref{fig:hwvspyr}.}\label{fig:fullvalidation}
\end{figure}

Table~\ref{tab:fullvalidation} and fig.~\ref{fig:fullvalidation} show
the resulting multiplicities, demonstrating good agreement. The two
samples are statistically independent for these quantities, since at
fixed collision energy the massless shower does not depend on the
orientation or light flavour of the hard event. The distributions of
figs.~\ref{fig:hwvspyr} and~\ref{fig:fullvalidation} are also
compatible. In both samples, four-momentum is conserved in every event
to within $\mathcal{O}(10^{-12})$~GeV. We emphasize that parton multiplicities and single-parton spectra depend on the shower cut-off and are used here as implementation diagnostics rather than infrared-safe observables.

\begin{table}[tbp]
\centering
\begin{tabular}{lcc}
\hline
Quantity & \texttt{Herwig 7.3.0} & \texttt{Pyresias}\\
\hline
Mean gluon count & $4.3166\pm0.0024$ & $4.3182\pm0.0024$\\
Mean quark + antiquark count & $2.24399\pm0.00070$ & $2.24519\pm0.00070$\\
Fraction with secondary pairs & $0.11510\pm0.00032$ & $0.11559\pm0.00032$\\
Mean summed gluon energy [GeV] & $59.722\pm0.039$ & $59.747\pm0.039$\\
\hline
\end{tabular}
\caption{Event means for the samples of fig.~\ref{fig:hwvspyr}, with their standard errors.}\label{tab:fullvalidation}
\end{table}

\subsection{Suggested Exercises}\label{sec:exercises}

\begin{enumerate}
\item At fixed coupling and fixed $z$ bounds, set $\Gamma(t)=\kappa/t$. Derive $\Delta(t_0,T)=(t_0/T)^\kappa$, generate first-emission scales, and check both the continuous spectrum and the fraction of lines with no emission. Their probabilities must sum to one as in eq.~\ref{eq:unitarity}. Figure~\ref{fig:sudakov} shows the same test with the running coupling and scale-dependent bounds.
\item Multiply a valid overestimate by two and adjust its primitive consistently. The accepted distribution should remain unchanged within sampling errors, while the veto efficiency changes. Explain why restarting from the original upper scale after a veto would fail this check.
\item Run the full event script with \texttt{--quark-only}, then with \texttt{--flavours 0}, and finally with its default five flavours. The first setting retains only quark-line radiation; the second adds $g\rightarrow gg$; the last also adds secondary quark pairs. Identify which multiplicities test each extension.
\item Check the sum of final-state four-momenta before and after the global recoil step. Vary the transverse-momentum cut-off and examine the gluon multiplicity and energy spectrum. Explain why their cut-off dependence is compatible with event-by-event energy conservation.
\item The record of accepted and vetoed trials allows a line to be reweighted to a modified shower without evolving it again~\cite{Bellm:2016voq,Mrenna:2016sih}. For each trial, let $p$ and $p'$ be its acceptance probability with the original and with a modified value of $\alpha_S(M_Z)$, keeping $\hat\alpha_S$ an upper bound for both. Multiply the weight of the line by $p'/p$ for each accepted trial and by $(1-p')/(1-p)$ for each vetoed one. Compare the reweighted first-emission distribution with one generated directly with the modified coupling.
\end{enumerate}

\section{Conclusions}\label{sec:conclusions}

Parton showers are an essential ingredient in Monte Carlo simulations of collider events and have applications beyond collider physics. They use splitting functions to generate collinear emissions, including those that are also soft. Large-angle soft emissions exhibit colour coherence and can be treated through angular ordering. In this article, we have motivated the parton shower algorithm via the Sudakov form factor, by considering multiple gluon emissions in an $e^+e^-$ collision that generates a quark--antiquark pair. We have presented the Sudakov veto algorithm, used to implement the treatment of QCD radiation in such a formalism, and verified that it reproduces the no-emission probability and first-emission distribution obtained by direct integration. We have also discussed how the ordering variable, the recoil, and the veto algorithm connect to the logarithmic accuracy and uncertainties of parton showers, and to the other aspects of event generation that interface with the parton shower.

We then presented the implementation as a tutorial designed to introduce the basic concepts step by step: from a single quark line to complete $e^+e^-\rightarrow q\bar q$ events, and then to the gluon branchings $g\rightarrow gg$ and $g\rightarrow q\bar q$ through competition between branching channels. With matching parton-level settings, the resulting events agree with the angular-ordered shower of the widely used \texttt{Herwig 7} general-purpose event generator within the statistical uncertainties of one million events. This validates the implementation within the stated massless, parton-level set-up; a precision prediction would also require matching, hadronization, and the other components of a general-purpose event generator. The accompanying code and exercises allow the reader to modify each of these ingredients and test the consequences.

\acknowledgments

We are grateful to Siddharth Sule for useful discussions and suggestions. A.P.\ acknowledges support from the National Science Foundation under Grant No.\ PHY 2210161.

\section*{Statements and declarations}

\paragraph{Funding.} A.P.\ acknowledges support from the National Science Foundation under Grant No.\ PHY 2210161.

\paragraph{Competing interests.} The author has no relevant financial or non-financial interests to disclose.

\paragraph{Data and code availability.} The \texttt{Pyresias} code, notebook, and hard-event inputs, together with the scripts and histogram data used for the figures of section~\ref{sec:results}, are publicly available~\cite{repo}.

\paragraph{Use of AI tools.} The author used AI
  assistants (OpenAI Codex Astra 6; Anthropic Claude Opus 5.5)
  during the preparation of this work to draft and revise parts of
  the text. The
  author edited all AI-assisted text and takes full responsibility for the content.


\appendix

\section{Angular Distribution for a Single Gluon Emission}\label{app:angular-distribution}

We derive the angular distribution of eq.~\ref{eq:ee-angular-singular}, including the non-singular term, from eq.~\ref{eq:ee-energy-distribution}. We work at tree level with massless quarks and gluons, in the centre-of-mass frame of the $e^+e^-$ collision. The outgoing quark, antiquark, and gluon have four-momenta $p_1$, $p_2$, and $p_3$, respectively, with $p_1+p_2+p_3=q$ and $q^\mu=(Q,\mathbf{0})$. As in the main text, $x_i=2E_i/Q$, $x_1+x_2+x_3=2$, and $\theta$ is the angle between the quark and the gluon. We retain the exact recoil kinematics throughout.

The invariant mass of the quark--gluon pair can be written in two ways:
\begin{equation}\label{eq:angular-invariant}
 (p_1+p_3)^2=(q-p_2)^2=Q^2(1-x_2)
 =\frac{Q^2}{2}x_1x_3(1-\cos\theta)\;.
\end{equation}
Using $x_2=2-x_3-x_1$, this gives:
\begin{equation}
 x_1\left[1-\frac{x_3}{2}(1-\cos\theta)\right]=1-x_3\;.
\end{equation}
For convenience, define $D=2-x_3(1-\cos\theta)$. The quark energy fractions are then:
\begin{equation}\label{eq:angular-energy-fractions}
 x_1=\frac{2(1-x_3)}{D}\;,
 \qquad x_2=2-x_3-\frac{2(1-x_3)}{D}\;.
\end{equation}
Since $x_2=2-x_3-x_1$, the absolute Jacobian for the change of variables is:
\begin{equation}\label{eq:angular-jacobian}
 \left|\frac{\partial(x_1,x_2)}{\partial(x_3,\cos\theta)}\right|
 =\left|\frac{\partial x_1}{\partial\cos\theta}\right|_{x_3}
 =\frac{2x_3(1-x_3)}{D^2}\;.
\end{equation}
We also obtain:
\begin{equation}\label{eq:angular-denominators}
 1-x_1=\frac{x_3(1+\cos\theta)}{D}\;,
 \qquad 1-x_2=\frac{x_3(1-x_3)(1-\cos\theta)}{D}\;.
\end{equation}
Substituting these relations and the Jacobian into eq.~\ref{eq:ee-energy-distribution} gives:
\begin{equation}\label{eq:angular-transformed}
 \begin{aligned}
 \frac{\mathrm{d}\sigma}{\mathrm{d}x_3\,\mathrm{d}\cos\theta}
 &=\sigma_0\frac{C_F\alpha_S}{2\pi}
 \frac{x_1^2+x_2^2}{(1-x_1)(1-x_2)}
 \frac{2x_3(1-x_3)}{D^2}\\
 &=\sigma_0\frac{C_F\alpha_S}{2\pi}
 \frac{2(x_1^2+x_2^2)}{x_3(1-\cos^2\theta)}\;.
 \end{aligned}
\end{equation}

To separate the singular terms, we use $x_1+x_2=2-x_3$ to write the numerator as:
\begin{equation}
 x_1^2+x_2^2=1+(1-x_3)^2-2(1-x_1)(1-x_2)\;.
\end{equation}
The last term cancels the angular denominator in eq.~\ref{eq:angular-transformed}, since:
\begin{equation}
 (1-x_1)(1-x_2)=\frac{x_3^2(1-x_3)(1-\cos^2\theta)}{D^2}\;.
\end{equation}
We therefore obtain the exact result:
\begin{equation}\label{eq:ee-angular-exact}
 \frac{\mathrm{d}\sigma}{\mathrm{d}x_3\,\mathrm{d}\cos\theta}
 =\sigma_0\frac{C_F\alpha_S}{2\pi}
 \left[\frac{2}{\sin^2\theta}\frac{1+(1-x_3)^2}{x_3}
 -\frac{4x_3(1-x_3)}{\left[2-x_3(1-\cos\theta)\right]^2}\right]\;.
\end{equation}
For fixed $0<x_3<1$, the second term in square brackets remains finite
as $\theta\rightarrow0$ or $\theta\rightarrow\pi$. It also behaves as
$-x_3+\mathcal O(x_3^2)$ in the soft-gluon limit,
$x_3\rightarrow0$. The first term thus contains the two collinear
poles and the soft-gluon singularity, while the second is the
non-singular term of eq.~\ref{eq:ee-angular-singular}. The result uses
the quark--gluon angle in the overall centre-of-mass frame and makes no assumption that the quark and antiquark are back-to-back.

\section{Proof of the Sudakov Veto Algorithm}\label{app:sudakovproof}

To prove the Sudakov veto algorithm,\footnote{This proof has been adapted from Ref.~\cite{Lonnblad:2012hz}.} described in section~\ref{sec:SudakovVeto}, consider the total probability of \textit{not} having an emission in $(t_0, t_\mathrm{max})$ after rejecting (i.e.\ vetoing) any number of trials:
\begin{equation}
\bar{P}_\mathrm{tot} = \sum_{n=0}^\infty \bar{P}_n\;,
\end{equation}
where $\bar{P}_n$ is the joint probability of rejecting exactly $n$ intermediate trials and reaching the cut-off without an accepted emission. For \textit{no} vetoed trials ($n=0$):

\begin{equation}
\bar{P}_0 = \hat{\Delta}(t_0, t_\mathrm{max})\;,
\end{equation}
where $\hat{\Delta}(t_0, t_\mathrm{max})$ is the underestimate of the true Sudakov factor, defined by eq.~\ref{eq:SudakovUnderestimate}.
The contribution with exactly one rejected trial and no accepted emission is:
\begin{equation}
\bar{P}_1 = \int_{t_0}^{t_\mathrm{max}} \mathrm{d} t \hat{\Gamma}(t) \hat{\Delta}(t,t_\mathrm{max}) \left[ 1 - \frac{\Gamma(t)}{\hat{\Gamma}(t)} \right] \hat{\Delta}(t_0, t)\;,
\end{equation}
where we have integrated over all trial scales $t$ that have been rejected, and the factor $\hat{\Gamma}(t) \hat{\Delta}(t,t_\mathrm{max})$ represents the probability of a first trial emission occurring at $t$, the factor $\left[ 1 - \frac{\Gamma(t)}{\hat{\Gamma}(t)} \right]$ represents the probability that we have rejected it, and $\hat{\Delta}(t_0, t)$ ensures that there are no further emissions between $t$ and $t_0$.

By using the property of the Sudakov form factor that $\hat{\Delta}(t_0, t) \hat{\Delta} (t,t_\mathrm{max}) = \hat{\Delta}(t_0,t_\mathrm{max})$, we can then write:

\begin{equation}
\bar{P}_1 = \hat{\Delta}(t_0, t_\mathrm{max})\int_{t_0}^{t_\mathrm{max}} \mathrm{d} t   \left[ \hat{\Gamma}(t) - \Gamma(t) \right] \;.
\end{equation}

For two vetoed emissions, the scales obey $t_0<t_2<t_1<t_{\mathrm{max}}$. Define $h(t)=\hat\Gamma(t)-\Gamma(t)$. The product of trial Sudakov factors is
\begin{equation}
 \hat\Delta(t_1,t_{\mathrm{max}})\hat\Delta(t_2,t_1)\hat\Delta(t_0,t_2)
 =\hat\Delta(t_0,t_{\mathrm{max}})\;.
\end{equation}
Consequently,
\begin{equation}
 \bar P_2=\hat\Delta(t_0,t_{\mathrm{max}})
 \int_{t_0}^{t_{\mathrm{max}}}\mathrm{d}t_1\,h(t_1)
 \int_{t_0}^{t_1}\mathrm{d}t_2\,h(t_2)
 =\frac{\hat\Delta(t_0,t_{\mathrm{max}})}{2!}
 \left[\int_{t_0}^{t_{\mathrm{max}}}\mathrm{d}t\,h(t)\right]^2\;.
\end{equation}
For $n$ vetoes, the ordered integration region similarly gives
\begin{equation}
 \bar P_n=\frac{\hat\Delta(t_0,t_{\mathrm{max}})}{n!}
 \left[\int_{t_0}^{t_{\mathrm{max}}}\mathrm{d}t\,
 (\hat\Gamma(t)-\Gamma(t))\right]^n\;.
\end{equation}
Summing over the number of vetoes yields
\begin{equation}
 \bar P_{\mathrm{tot}}=\hat\Delta(t_0,t_{\mathrm{max}})
 \exp\left[\int_{t_0}^{t_{\mathrm{max}}}\mathrm{d}t\,
 (\hat\Gamma(t)-\Gamma(t))\right]\;.
\end{equation}

Comparing with eq.~\ref{eq:SudakovUnderestimate}, we see that the exponential cancels the excess suppression in the trial Sudakov factor, giving
\begin{equation}
    \bar{P}_\mathrm{tot} = \Delta(t_0, t_\mathrm{max})\;,
\end{equation}
i.e.\ the original Sudakov form factor.

We have therefore shown that the SVA recovers the correct Sudakov form factor as the probability of no emission in $(t_0,t_\mathrm{max})$, as required. Since the argument holds for any lower limit, replacing $t_0$ by an arbitrary $t<t_\mathrm{max}$ shows that the probability of no accepted emission above $t$ is $\Delta(t,t_\mathrm{max})$. The first accepted emission is therefore distributed according to eq.~\ref{eq:pnextemission}.

\section{Rotations}\label{app:rotations}
To rotate a three-vector $\mathbf{a}$ onto the direction of a three-vector $\mathbf{b}$, we first construct the axis unit vector by taking their cross product: $\hat{\mathbf{k}} = \frac{\mathbf{a} \times \mathbf{b}}{|\mathbf{a} \times \mathbf{b}|}$. Then, the rotation is given by the Euler--Rodrigues rotation formula:
\begin{equation}
    \mathbf{R} = \mathbf{I} + (\sin \theta) \mathbf{K} + (1-\cos \theta) \mathbf{K}^2\;,
\end{equation}
where $\mathbf{I}$ is the identity matrix, $\theta$ is the angle between $\mathbf{a}$ and $\mathbf{b}$, and the matrix $\mathbf{K}$ takes the form:
\begin{equation}
\mathbf{K} = \begin{pmatrix}
0 & -\hat{k}_z & \hat{k}_y \\
\hat{k}_z & 0 & -\hat{k}_x \\
-\hat{k}_y & \hat{k}_x & 0
\end{pmatrix}\;,
\end{equation}
For parallel vectors, the rotation is the identity. For antiparallel vectors, choose any unit axis perpendicular to $\mathbf a$ and rotate by $\pi$; the cross-product formula for the axis is undefined in these two limits. For a generic non-zero pair, the rotated vector $\mathbf{a}'$ is given by $\mathbf{a}' = \mathbf{R} \mathbf{a}$.


\bibliographystyle{JHEP}
\bibliography{biblio.bib}

\end{document}